# Interpreting NMR dynamic parameters via the separation of reorientational motion in MD simulation

Albert A. Smith[1*]

[1]*Institute for Medical Physics and Biophysics, Leipzig University, Härtelstraße 16-18, 04107 Leipzig, Germany*

albert.smith-penzel@medizin.uni-leipzig.de

## Abstract

Reorientational dynamics—motion defined by changes in the direction of a vector or tensor—determine relaxation behavior in nuclear magnetic resonance (NMR). However, if multiple processes exist that result in reorientation, then analyzing the net effects becomes a complex task, so that one ideally would separate those motions to simplify analysis. The model-free and two-step approaches have established the separability of the total correlation function of reorientation motion into contributions from statistically independent motions. Separability has been used to justify the analysis of experimental relaxation rate constants by fitting data to a total correlation function resulting from the product of two or three individual correlation functions, each representing an independent motion. The resulting parameters are used to describe motion in the molecule, but if multiple internal motions are present, interpreting those parameters is not trivial. We suggest an alternative approach: quantitative and timescale-specific comparison of experiment and simulation, as previously established using the detector analysis. This is followed by separation of simulated correlation functions into independent motions, and timescale-specific parameterization of the results, such that one may determine how each motion contributes to experimental parameters. We establish protocols for the separation of the correlation function into components using coordinates from molecular dynamics simulation. Separation is achieved by defining a series of frames, where the frames iteratively split the total motion into contributions from motion within each frame and of each frame. Then timescale specific parameters (e.g. detector responses) describing the total motion may be interpreted in terms of the timescale-specific parameterization of the individual motions.

# 1   Introduction

How should we understand biomolecular function? It certainly helps to see the structure [1,2], but we know that the molecule exists not in one structure, but rather an ensemble of states that are sampled at thermal equilibrium [3,4] . So, have we understood the system if we know the complete ensemble of states? The transition between two low-energy states in the ensemble requires traversing many in-between states, and multiple pathways may exist for the transition to occur. Then, the ensemble alone omits this connectivity between individual structures. We could watch a movie of the transition between states, e.g. molecular dynamics (MD) simulations may further elucidate some dynamics processes, showing the pathways between states. However, simply watching the simulation is sometimes more confusing than clarifying. The overwhelming amount of motional information leaves one wondering what dynamics are connected and critical to function and which ones just happen to occur in the given trajectory. Even if we may separate the influence of individual motions, we still would like to verify the timescale and amplitudes of the simulated motions, noting that force field accuracy can still be improved [5,6], so that experimental verification by methods such as nuclear magnetic resonance (NMR) relaxation are critical [7–11].

With these challenges in mind, we realize just how attractive a combination of NMR relaxation and MD simulation is for characterizing complex dynamics. While MD provides a high level of detail, it often requires experimental validation. Relaxation of a given nucleus in NMR is usually dominated by the reorientational motion of one or several anisotropic interactions [12,13]. For example, backbone $^{15}$N relaxation is determined almost entirely by reorientational motion of the one-bond $^{1}$H–$^{15}$N dipole tensor and the $^{15}$N chemical shift anisotropy (CSA). Similarly, in specifically labeled methyl groups of protein side chains ($^{13}$CH$_1$D$_2$, [9,14–16]), $^{13}$C relaxation is determined by the dipole couplings to the $^{1}$H and $^{2}$H, as well as the $^{13}$C CSA (chemical exchange may also influence transverse relaxation for both cases [17]). What makes NMR particularly fascinating is that reorientation of a given interaction is a result of the aggregated influence of many separate modes of motion. A given motion may be local, influencing relaxation of one or only a few nuclei, but it also may be collective, with its motion projected onto multiple tensors and affecting the relaxation of many nuclei [10,18]. This manifests in the motion of a given interaction tensor as a multi-exponential decay of its correlation function. NMR relaxation is sensitive to this multi-exponential behavior, where a given relaxation measurement is dominated by components of the correlation function that have correlation times near the inverse of the eigenfrequencies of the spin-system ($\tau_c \sim 1/\omega$) [19,20]; therefore, a given experiment is timescale selective, and one may vary experiment ($T_1$, $T_{1\rho}$, NOE, etc.) and experimental conditions (external field, spin-lock field strength, MAS frequency, etc.) to change the eigenfrequencies, and therefore sample different timescales.

Then, is it possible to separate both the influences of multiple motions based only on NMR data? In certain cases, yes, the most well-known being that one may separate tumbling motion of a molecule in solution, which influences relaxation of all nuclei, from the net influence of all internal motions via the model-free or related approaches [21–28]. In this case, separability is brought about due to the presence of one large, well-defined motion with only a few parameters describing it, but manifesting on many nuclei (separating anisotropic tumbling requires knowing the average molecular structure, in order to correctly account for its influence on interaction tensors having different orientations). For isotropic tumbling, the model-free [25–27] and related two-step approaches [21–24,28] assume the correlation function for the rank-2 tensor reorientation to be either

$$C(t) = \underbrace{\exp(-t/\tau_M)}_{C_O(t)} \underbrace{\left[ S^2 + (1-S^2)\exp(-t/\tau) \right]}_{C_I(t)} \quad \text{model-free}$$

$$C(t) = \underbrace{S^2 \exp(-t/\tau_M)}_{C_O(t)} + \underbrace{(1-S^2)\exp(-t/\tau)}_{C_I(t)} \quad \text{two-step}$$

(1)

The correlation functions describe the correlation of an interaction tensor's orientation at some initial time ($\tau$) with its orientation at a later time ($t+\tau$). Then, for $t=0$, we see that $C(t) = 1$, that is, the orientation is fully correlated with itself, and then that correlation decays. Two causes of decay appear in this form: internal motions ($C_I(t)$), leading to an amplitude of decay of $(1-S^2)$ and time constant of decay given by $\tau$, and second influence from isotropic tumbling, which has an amplitude of decay of 1 (fully decaying to zero, since the final orientation has no correlation with the initial orientation), and correlation time $\tau_M$. Note that the two approaches become equivalent if the timescales of internal motion and of tumbling are well separated ($\tau_M \gg \tau$).

While such approaches are powerful, one still only separates tumbling from internal motion, and parameters $S^2$ and $\tau$ are the result of the collective effects of multiple internal motions. An extension of the model-free approach separates internal motion into fast and slow components, yielding internal parameters ($\tau_f$, $S_f^2$) and ($\tau_s$, $S_s^2$), respectively [29] (note that the theoretical basis for this extension did not appear in the original paper, but was later justified [28]). This *extended model-free* approach has also been adopted into solid-state NMR, where tumbling is no longer included, but two or three internal motions are fitted [30–33].

$$C(t) = \underbrace{\exp(-t/\tau_r)}_{C_o(t)} \underbrace{\left(S_f^2 + (1-S_f^2)\exp(-t/\tau_f)\right)\left(S_s^2 + (1-S_s^2)\exp(-t/\tau_s)\right)}_{C_l(t)} \quad \text{solution-state}$$

$$C(t) = \left(S_f^2 + (1-S_f^2)\exp(-t/\tau_f)\right)\left(S_s^2 + (1-S_s^2)\exp(-t/\tau_s)\right) \quad \text{solid-state}$$

(2)

In general, however, NMR relaxation of a given nucleus is the net effect of not just 1-3 motions but rather of many modes of motions, for which we cannot obtain sufficient experimental data to fully separate the influences of all independent motions. This has recently motivated us to develop the dynamics *detectors* approach of analyzing relaxation data, where we suppose that there are an arbitrary number of modes influencing the motion, and so to avoid biasing, we account for this in the assumed form of the correlation function such that it may contain an arbitrary number of decaying exponential terms [34,35]:

$$C(t) = S^2 + (1-S^2)\int_{-\infty}^{\infty} \theta(z)\exp\left(-t/(10^z \cdot 1\,\text{s})\right)dz \tag{3}$$

Then, $S^2$ is the generalized order parameter for all motion of a given tensor, and $\theta(z)$ describes how that motion is distributed as a function of the log-correlation time ($z = \log_{10}(\tau_c / 1\,\text{s})$). Without any assumptions about the form of $(1-S^2)\theta(z)$, clearly we cannot fully define its amplitude for all values of $z$. However, we may characterize the amplitude of motion for a given range of correlation times using detectors. This is characterized by a detector response, $\rho_n^{(\theta,S)}$.

$$\rho_n^{(\theta,S)} = (1-S^2)\int_{-\infty}^{\infty} \theta(z)\rho_n(z)dz \tag{4}$$

The detector sensitivity, $\rho_n(z)$, selects a range of correlation times as is illustrated in Fig. 1. From a given data set, we obtain multiple detector responses, each characterizing a different range of correlation times (although with some overlap), therefore yielding a timescale specific characterization of the dynamics. We have also adapted this approach to yield a model-free type separation of internal and tumbling motion [36,37].

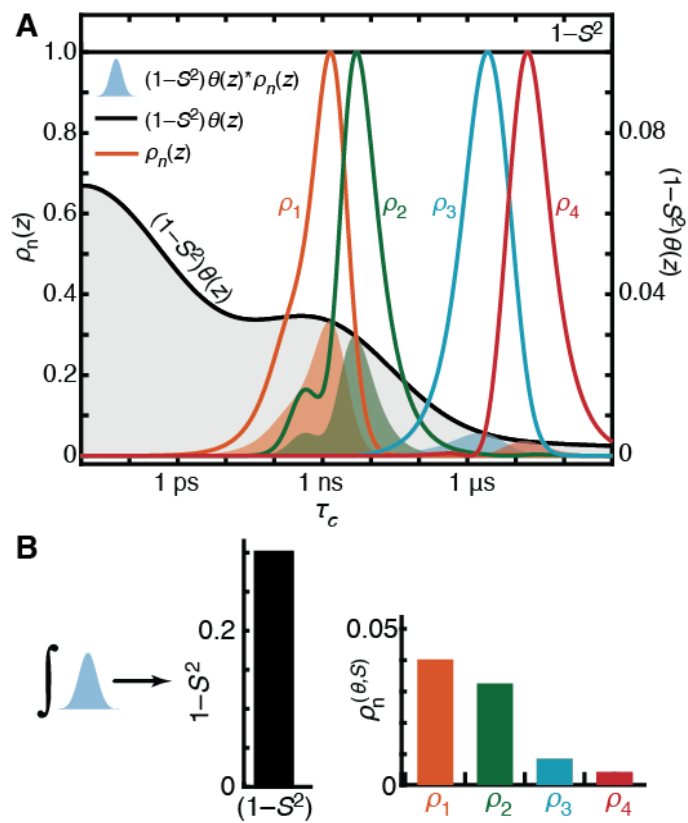

**Fig. 1.** Detector windows. If we assume that the correlation function of reorientational motion for a given NMR interaction is given by a distribution of correlation times, denoted $(1-S^2)\theta(z)$ (**A**, black line, right axis), then the integral of the distribution of motion is $(1-S^2)$, so that measuring a residual coupling to obtain $S$ can be used to characterize the total motion, without taking timescale into account. Detector responses, on the other hand, depend on their sensitivities, denoted $\rho_n(z)$ (**A**, colored lines). A detector response, $\rho_n^{(\theta,S)}$, depends only on the part of the distribution of correlation times that is found within the window defined by $\rho_n(z)$ (**A**, shaded areas), where detector responses are the integral of the shaded regions, with the integrals given in a bar plot in **B**. One may roughly interpret detector responses as the amplitude of reorientational motion around the center of the given detector sensitivity. See [35] for further details.

The detector responses separate motion by timescale, but not by the type of the motion, often yielding a still convoluted analysis of the dynamics. However, we may also characterize an MD simulation using detectors with approximately the same sensitivities, so that we have a quantitative comparison of timescale-specific motion in the experiment and simulation [10,38,39]. If reproduction of detector responses is reasonably good, we may perform further analysis on the MD simulation to determine the origin of motions in each detector window (i.e. at each timescale), thus connecting the experimental observables to explicit motions.

The question, however, is how do we go about separating the contributions of individual motions to the total reorientational dynamics of a tensor using an MD trajectory? In Fig. 2, we illustrate an example, where we note that the reorientational motion of an H–C bond (and therefore the H–C dipole tensor) in a methyl group of isoleucine is the result of methyl rotation, rotation around $\chi_1$ and $\chi_2$ angles, and motion of the Cα–Cβ bond (with additional motion due to fluctuations of bond angles away from their mean values). The total rotation may be decomposed into several steps corresponding to these contributions. Under certain circumstances, it is also possible to decompose the total correlation function into contributions from each rotation, such that

$$C(t) = C^1(t) \cdot C^2(t) \cdot \ldots \cdot C^N(t). \tag{5}$$

If this separation of the correlation function is achieved, we may analyze each individual MD-derived correlation function separately to determine the amplitude and timescale of each motion and how they contribute to the total reorientational dynamics of a given NMR tensor.

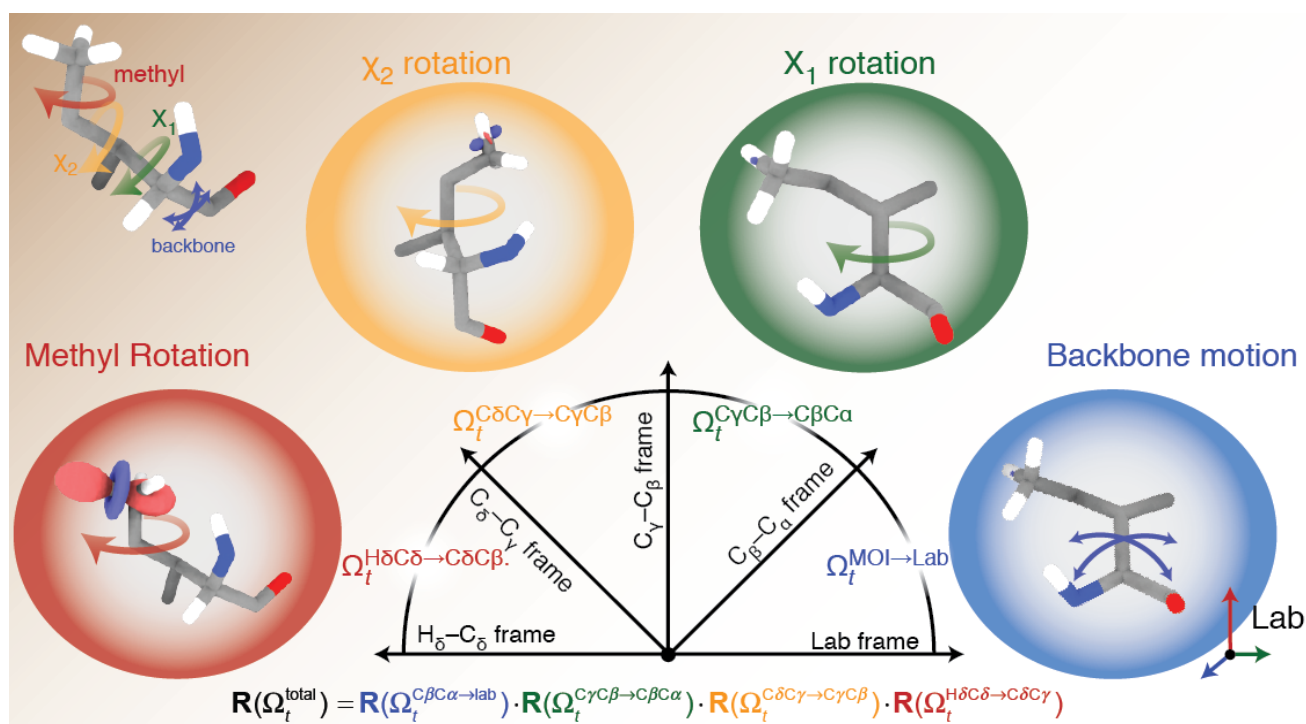

**Fig. 2.** Decomposing the total rotation of a methyl group of isoleucine into components. Here, we suppose that the total rotation of the H–C dipole tensor from its principle axis system to the lab frame is due to: 1) methyl rotation (red), defined as the rotation from the Hδ–Cδ frame to the Cδ–Cγ frame. 2) χ$_2$ rotation (orange), defined as the rotation from the Cδ–Cγ frame to the Cγ–Cβ frame. 3) χ$_1$ rotation (green), defined as the rotation from the Cγ–Cβ frame to the Cβ–Cα frame. 4) Backbone motion (blue), defined as rotation from the Cβ–Cα frame to the lab frame. Each frame shows the residual tensor resulting from all prior motions centered on the H–C bond. Note then, for the first motion, no attenuation has occurred yet. For the last motion, the tensor is nearly invisible due to attenuation from the previous motions. Figure concept from [40], molecules with tensors created with ChimeraX [41].

In this study, we will start from the model-free and related theories [21–28], and review the conditions under which separation of motion is possible, such that we obtain the total correlation function as a product of correlation functions of individual motions. We will then demonstrate how to obtain the individual correlation functions from an MD trajectory, and use the correlation functions to investigate the influence that individual motions have on simulated and, by extension, experimental detector responses. Therefore, we may separate and characterize complex dynamics in simulation by both type and timescale. We refer to the proposed approach as <u>R</u>e-<u>O</u>rientational dynamics in <u>M</u>D <u>a</u>nalyzed for <u>N</u>MR <u>c</u>orrelation function dis<u>e</u>ntanglement, or ROMANCE for short. Note that several related approaches exist. In particular, Salvi, Abyzov, and Blackledge have achieved the desired separation, but only for the case of backbone dynamics in an intrinsically disordered protein (IDP), separated into librations, φ,ψ motion, and peptide plane tumbling [42]. Skrynnikov and coworkers use frame alignment to remove specific motions to assess correlation between residues in an IDP [43], Vogel et al. use frames to analyze backbone H–N motions within α-helices of $Y_1$ GPCR and motion of the α-helices themselves but do not determine how motion of the helix influences the correlation function of the H–N bond [44], and Prompers and Brüschweiler separate the correlation function into a sum (rather than a product) of contributions from eigenmodes of the rank-2 angular correlation matrix [45,46]. We have recently published an application of the ROMANCE method detailed here, using it to separate different types of motion in a POPC lipid membrane [38].

## 2   Theory

Reorientational motion results from internal dynamics of a molecule, or from the re-orientation of the molecule itself, and manifests in NMR relaxation rate constants. For example, NMR relaxation of a given spin-1/2 nucleus (X) is usually dominated by reorientation of dipolar interactions with bonded hydrogens (H) and by the chemical shift anisotropy (CSA) tensor of that nucleus. For spins greater than 1/2, the quadrupolar coupling is usually dominant, and for transverse relaxation, changes in the isotropic chemical shift also contribute to the relaxation. The results here are easily applied to quadrupolar relaxation, but relaxation by change in chemical shift is not the direct result of reorientational motion, so we will not consider it in this study.

To begin, we denote the direction of the principle component of an interaction tensor with a vector, $\mathbf{v}_Z(\tau)$, which is a normalized, time-dependent vector with components $\mathbf{v}_Z(\tau) = [x_Z(\tau), y_Z(\tau), z_Z(\tau)]$. Usually, this vector will be parallel with some bond in the molecule (for example, an H–X dipole coupling's principle component is along the H–X bond), although may have a more complex dependence on the local geometry, for example, the principle component of the $^{15}$N CSA in the protein backbone lies ~23° away from the H–N bond. Later, we will define a full axis system for the interaction, which will have *x*-, *y*-, and *z*-axes, but for now we only require the *z*-axis. Then, time-dependence of the tensor reorientation can be expressed via rotation matrices, such that

$$\mathbf{v}_Z(t+\tau) = \mathbf{R}_{ZYZ}(\Omega_{\tau,t+\tau}^{LF}) \cdot \mathbf{v}_Z(\tau). \tag{6}$$

$\mathbf{R}(\Omega_{\tau,t+\tau}^{LF})$ is a rotation matrix with Euler angles $\Omega_{\tau,t+\tau}^{LF} = [\alpha_{\tau,t+\tau}^{LF}, \beta_{\tau,t+\tau}^{LF}, \gamma_{\tau,t+\tau}^{LF}]$. The superscript denotes that the Euler angles are defined in the lab frame (LF), and the subscripts denotes that this is the rotation of the vector from its orientation at time $\tau$ to its orientation at time $t+\tau$. We will follow the usual zyz-convention for rotation matrices, so that this matrix may be expanded as

$$\begin{aligned}\mathbf{R}_Z(\Omega_{\tau,t+\tau}^{LF}) &= \mathbf{R}_Z(\gamma_{\tau,t+\tau}^{LF}) \cdot \mathbf{R}_Y(\beta_{\tau,t+\tau}^{LF}) \cdot \mathbf{R}_Z(\alpha_{\tau,t+\tau}^{LF}) \\ &= \begin{bmatrix} \cos\gamma_{\tau,t+\tau}^{LF} & -\sin\gamma_{\tau,t+\tau}^{LF} & 0 \\ \sin\gamma_{\tau,t+\tau}^{LF} & \cos\gamma_{\tau,t+\tau}^{LF} & 0 \\ 0 & 0 & 1 \end{bmatrix} \cdot \begin{bmatrix} \cos\beta_{\tau,t+\tau}^{LF} & 0 & \sin\beta_{\tau,t+\tau}^{LF} \\ 0 & 1 & 0 \\ -\sin\beta_{\tau,t+\tau}^{LF} & 0 & \cos\beta_{\tau,t+\tau}^{LF} \end{bmatrix} \cdot \begin{bmatrix} \cos\alpha_{\tau,t+\tau}^{LF} & -\sin\alpha_{\tau,t+\tau}^{LF} & 0 \\ \sin\alpha_{\tau,t+\tau}^{LF} & \cos\alpha_{\tau,t+\tau}^{LF} & 0 \\ 0 & 0 & 1 \end{bmatrix}.\end{aligned} \tag{7}$$

Then, for any rotation, it is possible to separate the total rotation into several steps, due to closure of the rotation group, that is,

$$\mathbf{R}_{ZYZ}(\Omega_{\tau,t+\tau}) = \mathbf{R}_{ZYZ}(\Omega_{\tau,t+\tau}^{N-1,N}) \cdot \ldots \cdot \mathbf{R}_{ZYZ}(\Omega_{\tau,t+\tau}^{3,2}) \cdot \mathbf{R}_{ZYZ}(\Omega_{\tau,t+\tau}^{2,1}). \tag{8}$$

Each rotation matrix represents motion of the vector due to a different contribution to the overall motion. For example, if the dipolar tensor corresponds to an H–C bond located in a methyl group of an isoleucine, then the first rotation matrix ($\mathbf{R}_{ZYZ}(\Omega_{\tau,t+\tau}^{2,1})$) would describe primarily methyl rotation (around Cγ–Cδ), the second matrix would correspond to rotation about Cβ–Cγ, the third matrix would correspond to rotation about Cα–Cβ, and finally the last rotation matrix could correspond to reorientation of the Cα–Cβ due to other motion such as motion of the backbone. This is illustrated in Fig. 2.

In order to calculate relaxation behavior from a trajectory, we need the relevant rank-2 tensor correlation function, $C(t)$, where the Fourier transform of $C(t)$ yields the spectral density, $J(\omega)$. From $J(\omega)$, we can calculate various experimental relaxation rate constants. For example,

$$\begin{aligned} J(\omega) &= \frac{2}{5} \int_0^\infty C(t) \exp(-i\omega t) dt \\ R_1 &= \left(\frac{\delta_{HC}}{4}\right)^2 \left(J(\omega_H - \omega_C) + 3J(\omega_C) + 6J(\omega_H + \omega_C)\right) + \frac{3}{4}(\omega_C \sigma_{zz})^2 J(\omega_C) \end{aligned} \tag{9}$$

yields the $^{13}$C $R_1$ relaxation rate for a two-spin, $^1$H–$^{13}$C system, with dipole coupling $\delta_{HC}$, CSA with zz-component $\sigma_{zz}$, and Larmor frequencies $\omega_H$ and $\omega_C$ for the $^1$H and $^{13}$C nuclei, respectively. Thus, we may connect experiment to simulation via the correlation function, since it may both be characterized with NMR and calculated via MD. In this study, we will focus on construction of the correlation function itself from a trajectory, and how to separate the total correlation function into contributions of individual motions.

**Table 1.** Notation

| Term | Description |
|---|---|
| $\mathbf{v}_Z(\tau)$ | The direction of the principle component of an NMR interaction tensor. Without superscript, this vector is in the lab frame. |
| $\mathbf{v}_{XZ}(\tau)$ | Vector defining the *xz*-plane of the NMR interaction. Used to calculate $\Omega_\tau^{\mathbf{v}}$, which then itself may be used to obtain $\mathbf{v}_X(\tau)$ and $\mathbf{v}_Y(\tau)$. |
| $\mathbf{v}_X(\tau), \mathbf{v}_Y(\tau)$ | Direction of the *x*- or *y*-axis of the frame defined by the NMR interaction (lab frame). |
| $\mathbf{v}_Z^{\mathbf{v}}(t+\tau)$ | Direction of the NMR interaction tensor at time $t+\tau$, but represented in the frame of the interaction at time $\tau$ (that is, $\mathbf{v}_Z^{\mathbf{v}}(\tau) = [0,0,1]'$, since it is in its own frame). |
| $\mathbf{v}_{XZ}^{\mathbf{v}}(t+\tau)$ | Vector defining the *xz*-plane of the NMR interaction, given in the frame of the interaction at time $\tau$. |
| $v_Z^f(\tau), v_{XZ}^f(\tau)$ | Vectors defining the *z*-axis and *xz*-plane of a frame, *f*. |
| $v_X^f(\tau), v_Y^f(\tau)$ | Vectors defining the *x*- and *y*-axes of some frame *f*. May be obtained from $\Omega_\tau^f$. |
| $\mathbf{v}_Z^{-F}(\tau), v_Z^{f-F}(\tau), \ldots$ | Vector where motion of frame *F* has been removed, via alignment of that frame. Obtained by calculating $v_Z^{f-F}(\tau) = \mathbf{R}_{ZYZ}^{-1}(\Omega_\tau^F) \cdot v_Z^f(\tau)$ (similarly for other vectors). |
| $\Omega_\tau^{\mathbf{v}} = \{\alpha_\tau^{\mathbf{v}}, \beta_\tau^{\mathbf{v}}, \gamma_\tau^{\mathbf{v}}\}$ | Euler angles yielding the direction of the NMR interaction at time $\tau$ in the lab frame. |
| $\Omega_\tau^f$ | Euler angles yielding the direction of the frame *f* at time $\tau$ in the lab frame. |
| $\Omega_\tau^{\mathbf{v}-f}$ | Euler angles yielding the direction of the NMR interaction at time $\tau$ in frame *f* (motion of *f* removed). |
| $\Omega_{\tau,t+\tau}^{\mathbf{v}}$ | Euler angles rotating the NMR interaction from its direction at time $\tau$ to its direction at time $t+\tau$, *in the frame of the NMR interaction at time $\tau$*. That is, $\mathbf{v}_Z^{\mathbf{v}}(t+\tau) = \mathbf{R}_{ZYZ}(\Omega_{\tau,t+\tau}^{\mathbf{v}}) \cdot [0,0,1]'$, since $\mathbf{v}_Z^{\mathbf{v}}(\tau) = [0,0,1]'$. |
| $\Omega_{\tau,t+\tau}^{\mathbf{v}-f}$ | Euler angles rotating the NMR interaction from its $\tau$ to its direction at time $t+\tau$, *in the frame of the NMR interaction at time $\tau$*, where motion of frame *f* has been removed. |
| $\Omega_{\tau,t+\tau}^{\mathbf{v}:f}$ | Euler angles rotating the NMR interaction from its $\tau$ to its direction at time $t+\tau$, *in the frame of the NMR interaction at time $\tau$*, where that motion is due to the motion of *f*. |
| $\Omega_{\tau,t+\tau}^{\mathbf{v}:f-F}$ | Similar to $\Omega_{\tau,t+\tau}^{\mathbf{v}:f}$, except that the motion of another frame *F* has been removed before calculating motion due to frame *f*. |
| $C(t)$ | Correlation function resulting from the total reorientational motion of some interaction tensor. |
| $C^{\mathbf{v}-f}(t)$ | Correlation function resulting from all motion within frame *f* (motion of frame *f* removed). |
| $C^{\mathbf{v}f-F}(t)$ | Correlation function resulting from motion of frame *f*, with motion of frame *F* removed. |
| $C_{0p}^{\mathbf{v}-f}(t)$ | Shorthand for $\left\langle D_{0p}^2(\Omega_{\tau,t+\tau}^{\mathbf{v}-f}) \right\rangle_\tau$. $C_{00}^{\mathbf{v}-f}(t)$ is equal to $C^{\mathbf{v}-f}(\tau)$ |
| $C_{p0}^{\mathbf{v}f-F}(t)$ | Shorthand for $\left\langle D_{p0}^2(\Omega_{\tau,t+\tau}^{\mathbf{v}f-F}) \right\rangle_\tau$. Terms $C_{p0}^{\mathbf{v}f-F}(t)$ are used to construct $C^{\mathbf{v}f-F}(t)$ (section 2.2.2) |
| $A_p^{\mathbf{v}-f}$ | Equilibrium values of $C_{0p}^{\mathbf{v}-f}(t)$, such that $A_p^{\mathbf{v}-f} = \lim_{t\to\infty}\left\langle D_{0p}^2(\Omega_{\tau,t+\tau}^{\mathbf{v}-f}) \right\rangle_\tau = \lim_{t\to\infty} C_{0p}^{\mathbf{v}-f}(t)$. |
| $A_p^{-f}$ | Averaged tensor over all times, $\tau$, where $A_p^{-f} = \left\langle D_{0p}^2(\Omega_\tau^{\mathbf{v}-f}) \right\rangle_\tau$ (used for determining symmetry frame direction). |

## 2.1 Correlation function of total motion

The relevant correlation function resulting from all reorientational motions acting an NMR interaction tensor, which may be used to determine relation rate constants in NMR is calculated as

$$C(t) = \left\langle D_{00}^2(\Omega_{\tau,t+\tau}^v) \right\rangle_\tau = \left\langle \frac{3\cos^2 \beta_{\tau,t+\tau}^v - 1}{2} \right\rangle_\tau = \left\langle \frac{3(\mathbf{v}_Z(\tau) \cdot \mathbf{v}_Z(t+\tau))^2 - 1}{2} \right\rangle_\tau . \tag{10}$$

We take a moment to discuss the equation along with its notation. The correlation function of the total motion, $C(t)$, is obtained by calculating the average over all pairs of time points in a trajectory that are separated by $t$ (average over $\tau$), where this average is indicated by brackets, $\langle \ \rangle_\tau$. Then, for each pair of time points, we calculate an element of the Wigner rotation matrix, $D_{m',m}^2(\Omega)$, where $m'$ indicates the component of the initial spherical tensor that is transformed to component $m$ of a new spherical tensor by rotation with Euler angles $\Omega = \{\alpha, \beta, \gamma\}$. In eq. (10), we only require the $m'$=0, $m$=0 element (0,0), but later will need other elements of the matrix (matrix elements required for this study are given in Table 2). For this case, then we only require the angle $\beta_{\tau,t+\tau}^v$, for which the cosine may be obtained from the dot product of normalized vectors $\mathbf{v}(\tau)$ and $\mathbf{v}(t+\tau)$.

Then, the Euler angles required to calculate the correlation function of the total motion, denoted $\Omega_{\tau,t+\tau}^v$, rotate the vector from its orientation at time $\tau$ to its orientation at time $t+\tau$. A critical point: the rotation is performed in the frame of the vector at time $\tau$, not in the lab frame, as denoted by the superscript $\mathbf{v}$ ($\Omega_{\tau,t+\tau}^v \neq \Omega_{\tau,t+\tau}^{LF}$). If the calculation is performed in the lab frame, an additional factor of 1/5 appears in the correlation function, resulting from the orientational averaging of the sample either from tumbling in solution, or from the powder average in solid-state NMR (regardless of the frame choice, the factor of 1/5 must appear at some point; we insert it into the spectral density in eq. (9)). Then, $\mathbf{v}_Z^v(t+\tau)$ is defined in the frame of the vector at time $\tau$. We may calculate $\mathbf{v}_Z^v(t+\tau)$ from the Euler angles $\Omega_{\tau,t+\tau}^v$, according to

$$\mathbf{v}_Z^v(t+\tau) = \mathbf{R}(\Omega_{\tau,t+\tau}^v) \cdot \mathbf{v}_Z^v(\tau) = \mathbf{R}(\Omega_{\tau,t+\tau}^v) \cdot \begin{bmatrix} 0 \\ 0 \\ 1 \end{bmatrix} . \tag{11}$$

Since we are in the frame of the vector at time $\tau$, the initial vector must lie along the z-axis. Note that for brevity, we do not write $\mathbf{v}(\tau)$ in the superscript, which would fully specify that the frame is defined by the vector at time $\tau$ (as opposed to some other time), but shorten this to $\mathbf{v}$. An example of Euler angles rotating from $\mathbf{v}_z^\mathbf{v}(\tau)$ to $\mathbf{v}_z^\mathbf{v}(t+\tau)$ in the lab frame and the frame of $\mathbf{v}(\tau)$ is illustrated in Fig. 3.

**Table 2. Wigner Rotation Matrix Elements**

| p | -2 | -1 | 0 | 1 | 2 |
|---|---|---|---|---|---|
| $D^2_{0p}(\alpha,\beta,\gamma)$ | $e^{2i\gamma}\sqrt{\frac{3}{8}}\sin^2\beta$ | $-e^{i\gamma}\sqrt{\frac{3}{8}}\sin(2\beta)$ | $\frac{3\cos^2\beta-1}{2}$ | $e^{-i\gamma}\sqrt{\frac{3}{8}}\sin(2\beta)$ | $e^{-2i\gamma}\sqrt{\frac{3}{8}}\sin^2\beta$ |
| Real | $\sqrt{\frac{3}{8}}\left((c\gamma s\beta)^2-(s\gamma s\beta)^2\right)$ | $-\sqrt{\frac{3}{2}}(c\gamma s\beta)c\beta$ | $-\frac{1}{2}+\frac{3}{2}(c\beta)^2$ | $\sqrt{\frac{3}{2}}(c\gamma s\beta)c\beta$ | $\sqrt{\frac{3}{8}}\left((c\gamma s\beta)^2-(s\gamma s\beta)^2\right)$ |
| Imaginary | $\sqrt{\frac{3}{2}}(s\gamma s\beta)(c\gamma s\beta)$ | $-\sqrt{\frac{3}{2}}(s\gamma s\beta)c\beta$ | 0 | $-\sqrt{\frac{3}{2}}(s\gamma s\beta)c\beta$ | $-\sqrt{\frac{3}{2}}(s\gamma s\beta)(c\gamma s\beta)$ |
| $D^2_{p0}(\alpha,\beta,\gamma)$ | $e^{2i\alpha}\sqrt{\frac{3}{8}}\sin^2\beta$ | $e^{i\alpha}\sqrt{\frac{3}{8}}\sin(2\beta)$ | $\frac{3\cos^2\beta-1}{2}$ | $-e^{-i\alpha}\sqrt{\frac{3}{8}}\sin(2\beta)$ | $e^{-2i\alpha}\sqrt{\frac{3}{8}}\sin^2\beta$ |
| Real | $\sqrt{\frac{3}{8}}\left((c\alpha s\beta)^2-(s\alpha s\beta)^2\right)$ | $\sqrt{\frac{3}{2}}(c\alpha s\beta)c\beta$ | $-\frac{1}{2}+\frac{3}{2}(c\beta)^2$ | $-\sqrt{\frac{3}{2}}(c\alpha s\beta)c\beta$ | $\sqrt{\frac{3}{8}}\left((c\alpha s\beta)^2-(s\alpha s\beta)^2\right)$ |
| Imaginary | $\sqrt{\frac{3}{2}}(s\alpha s\beta)(c\alpha s\beta)$ | $-\sqrt{\frac{3}{2}}(s\alpha s\beta)c\beta$ | 0 | $\sqrt{\frac{3}{2}}(s\alpha s\beta)c\beta$ | $-\sqrt{\frac{3}{2}}(s\alpha s\beta)(c\alpha s\beta)$ |

*Each term is separated into real and imaginary components, and rearranged such that the required terms may be easily obtained from eq. (A5) or (A10).

Eq. (10) may be used to find the orientionally-averaged relaxation rate constants (e.g. eq. (9)). In general, the motion acting on a given tensor will result in different relaxation rate constants depending on the relative orientation of the tensor, the direction of the motion, and the direction of the external magnetic field (which determines the eigenstates of the spin system). However, for this study, we focus on calculating correlation functions that are orientationally averaged (averaged over the direction of the magnetic field). These are relevant in solution-state NMR, where molecular tumbling results in the measurement of the averaged rate constants. In a powder sample in solid-state NMR, angular dependence manifests as multi-exponential relaxation behavior. In case magic-angle spinning (MAS) is used, this orientational dependence is partially averaged. However, it is technically difficult to extract this orientational dependence, especially under MAS, so usually one extracts only the orientationally-averaged rate constant from experimental data (caution must be taken to correctly extract the averaged rate constants, see [47] for further discussion).

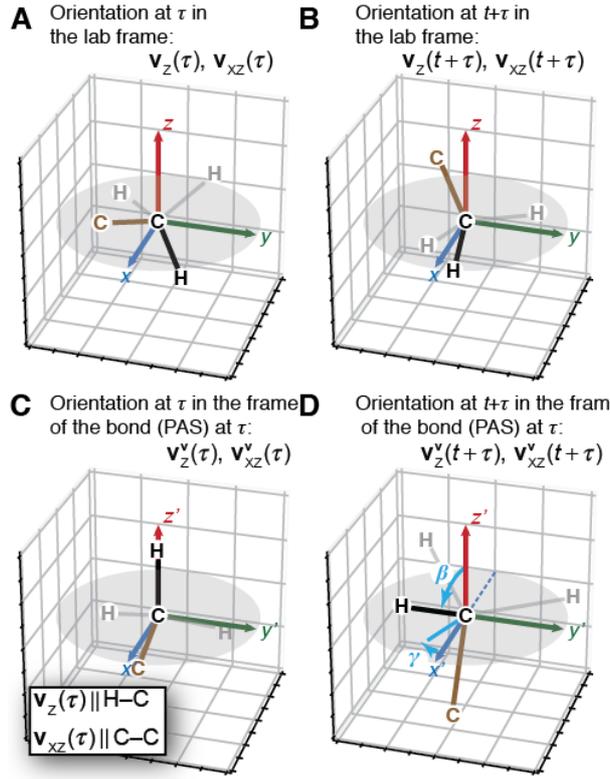

**Fig. 3.** Euler angles for transforming a bond vector at time $t+\tau$ in its PAS to the PAS of the bond vector at time $\tau$. We take a methyl group as an example, where the direction of the H–C bond in black/bold defines the direction of the z-axis of the frame of the interaction, $\mathbf{v}_z(\tau)$, and the C–C bond (brown) defines the xz-plane, $\mathbf{v}_{xz}(\tau)$. **A** illustrates an H–C bond (black, bold) as part of a methyl group at time $\tau$, and **B** illustrates the H–C bond at time $t+\tau$, both in the lab frame. **C** and **D** are illustrations of the bond transformed into the PAS of the bond at time $\tau$. Then, in **C**, the H–C bond lies along the z-axis, since it is in its own PAS. Furthermore, we define the PAS such that the C–C bond lies in the positive xz-plane. **D** shows the bond at time $t+\tau$, along with the $\beta$ and $\gamma$ Euler angles required to rotate from the z-axis to its position in the frame of the bond at time $\tau$ (that is, from its PAS at $t+\tau$ to its PAS at time $\tau$). Rotation by $\alpha$ about the z-axis (not shown) would result in bringing the C–C bond into the xz-plane, but would not influence the H–C bond, since it lies along z.

By performing calculations in the frame of $\mathbf{v}(\tau)$, we obtain the orientationally averaged correlation function directly. However, this is not the only approach to obtain the correlation function. Instead, we may express the correlation function as a sum of terms calculated in the lab frame [28]. In this case, the correlation function takes the following form:

$$C(t) = \sum_{m=-2}^{2} \left\langle V_m^{LF*}(\tau) V_m^{LF}(t+\tau) \right\rangle_\tau = 5 \left\langle V_0^{LF}(\tau) V_0^{LF}(t+\tau) \right\rangle_\tau. \qquad (12)$$

Here, the $V_m^{LF}(\tau)$ are components of the spherical tensors as a function of time. The correlation function to the left is given as a sum over the 5 components (m=-2,…,2), and is valid in the absence of molecular tumbling where different orientations of a given motion yield different relaxation behavior. If molecular tumbling is present, sampling all orientations equally, then the two functions become equal. For analysis of MD simulation, we cannot rely on the simulation sampling all orientations equally (this would take too much simulation time), so that we opt to use the correlation function in the form given in eq. (10). We point this out, because the methods of motional separation found in sections 2.2.1 and 2.2.3 are rigorously derived elsewhere [28], but in the lab frame. Therefore we revisit these here noting the change in frame, with explicit calculation of the individual correlation functions from an MD trajectory as our goal.

## 2.2 Separating two motions

Eq. (10) is the correlation function resulting from all motion causing reorientation of an interaction tensor. However, if we have two (or more) motions defined by separate rotations of the bond, can we separate the motion into its components? Let us suppose that we may define some intermediate frame, *f*, based on coordinates of a trajectory. Then, an interaction tensor, with time-dependent orientation defined by vector $\mathbf{v}_z(\tau)$, will reorient due to motion of the vector within frame *f*, and due to motion of frame *f* itself. That is,

$$\begin{aligned}\mathbf{v}^{\mathbf{v}}(t+\tau) &= \mathbf{R}(\Omega^{\mathbf{v}}_{\tau,t+\tau}) \cdot \mathbf{v}^{\mathbf{v}}(\tau) \\ &= \mathbf{R}(\Omega^{\mathbf{v}:f}_{\tau,t+\tau}) \cdot \mathbf{R}(\Omega^{\mathbf{v}-f}_{\tau,t+\tau}) \cdot \mathbf{v}^{\mathbf{v}}(\tau)\end{aligned} \quad (13)$$

The superscript **v** indicates that we are in the frame of the vector at time $\tau$. The superscript **v**-*f* indicates reorientation of the vector where motion of frame *f* is removed (motion within frame *f*). The superscript **v**:*f* indicates reorientation of the vector due to motion of frame *f*. The product of these two rotations should yield the total rotation as in (13). Then, it is possible to rewrite eq. (10), splitting the total rotation into contributions from the separated motions.

$$C(t) = \left\langle D^2_{00}(\Omega^{\mathbf{v}}_{\tau,t+\tau}) \right\rangle_\tau = \sum_{p=-2}^{2} \left\langle D^2_{0p}(\Omega^{\mathbf{v}:f}_{\tau,t+\tau}) D^2_{p0}(\Omega^{\mathbf{v}-f}_{\tau,t+\tau}) \right\rangle_\tau. \quad (14)$$

As has been previously shown [21–28], it may be possible to separate contributions from motion inside frame *f* and due to frame *f*, if those two motions are statistically independent, that is if

$$\left\langle D^2_{p0}(\Omega^{\mathbf{v}:f}_{\tau,t+\tau}) D^2_{0p}(\Omega^{\mathbf{v}-f}_{\tau,t+\tau}) \right\rangle_\tau = \left\langle D^2_{p0}(\Omega^{\mathbf{v}:f}_{\tau,t+\tau}) \right\rangle_\tau \left\langle D^2_{0p}(\Omega^{\mathbf{v}-f}_{\tau,t+\tau}) \right\rangle_\tau$$
then
$$C(t) = \sum_{p=-2}^{2} \left\langle D^2_{p0}(\Omega^{\mathbf{v}:f}_{\tau,t+\tau}) D^2_{0p}(\Omega^{\mathbf{v}-f}_{\tau,t+\tau}) \right\rangle_\tau = \sum_{p=-2}^{2} \left\langle D^2_{p0}(\Omega^{\mathbf{v}:f}_{\tau,t+\tau}) \right\rangle_\tau \left\langle D^2_{0p}(\Omega^{\mathbf{v}-f}_{\tau,t+\tau}) \right\rangle_\tau \quad (15)$$

Note that in general,

$$\left\langle D^2_{p0}(\Omega^{\mathbf{v}:f}_{\tau,t+\tau}) D^2_{0p}(\Omega^{\mathbf{v}-f}_{\tau,t+\tau}) \right\rangle_\tau = \left\langle D^2_{p0}(\Omega^{\mathbf{v}:f}_{\tau,t+\tau}) \right\rangle_\tau \left\langle D^2_{0p}(\Omega^{\mathbf{v}-f}_{\tau,t+\tau}) \right\rangle_\tau + \dots$$
$$\underbrace{\left\langle \left(D^2_{p0}(\Omega^{\mathbf{v}:f}_{\tau,t+\tau}) - \left\langle D^2_{p0}(\Omega^{\mathbf{v}:f}_{\tau,t+\tau}) \right\rangle_\tau\right) \left(D^2_{0p}(\Omega^{\mathbf{v}-f}_{\tau,t+\tau}) - \left\langle D^2_{0p}(\Omega^{\mathbf{v}-f}_{\tau,t+\tau}) \right\rangle_\tau\right) \right\rangle}_{\text{cov}\left(D^2_{p0}(\Omega^{\mathbf{v}:f}_{\tau,t+\tau}), D^2_{0p}(\Omega^{\mathbf{v}-f}_{\tau,t+\tau})\right)}, \quad (16)$$

so that statistical independence occurs when the covariance of the two terms vanishes.

Further assumptions are required to express $C(t)$ as the product of two correlation functions; in this form, terms resulting from motion in frame $f$ and terms resulting from motion of frame $f$ are within the sum, and so cannot be fully separated as two independent correlation functions. A few approaches to separate the correlation functions exist; we start with the classical model free approach, which assumes isotropic tumbling of the molecule in solution [25,26].

*2.2.1 Isotropic tumbling (model-free)*

In this case, we assume that the motion of the intermediate frame, $f$, is an overall tumbling motion. If the tumbling motion is isotropic, which is the case if the molecule is sufficiently spherical, then

$$\left\langle D^2_{p0}(\Omega^{v:f}_{\tau,t+\tau})\right\rangle_\tau = \left\langle \exp(-ip\alpha^{v:f}_{\tau,t+\tau})d^2_{p0}(\beta^{v:f}_{\tau,t+\tau})\right\rangle_\tau = \delta_{0,p}\left\langle d^2_{00}(\beta^{v:f}_{\tau,t+\tau})\right\rangle_\tau. \tag{17}$$

The symmetry of the motion means that while the distribution of $\beta$ angles depends on $t$, all $\alpha$ angles are equally likely. This causes the terms containing $\exp(-ip\alpha^{v:f}_{\tau,t+\tau})$ to average to zero, except for the case that $p=0$, where the exponential equals 1. Then, eq. (15) becomes

$$C(t) = \sum_{p=-2}^{2} \delta_{0,p} \left\langle d^2_{00}(\beta^{v:f}_{\tau,t+\tau})\right\rangle_\tau \left\langle D^2_{p0}(\Omega^{v-f}_{\tau,t+\tau})\right\rangle_\tau = \underbrace{\left\langle d^2_{00}(\beta^{v:f}_{\tau,t+\tau})\right\rangle_\tau}_{C^{v:f}(t)} \underbrace{\left\langle D^2_{00}(\Omega^{v-f}_{\tau,t+\tau})\right\rangle_\tau}_{C^{v-f}(t)}.$$
$$= C^{v:f}(t)C^{v-f}(t) \tag{18}$$

We denote the correlation function resulting from motion within frame $f$ as $C^{v-f}(t)$, and the function resulting from motion of frame $f$ as $C^{v:f}(t)$, where the product of these two functions yields $C(t)$. We note that $C^{v-f}(t)$ equilibrates at some final value, $\lim_{t\to\infty}\left\langle D^2_{00}(\Omega^{v-f}_{\tau,t+\tau})\right\rangle_\tau$, which is the definition of the familiar generalized order parameter, $S^2$. $S^2$ is related to the amplitude of the internal reorientational motion (motion within frame $f$), although that relationship depends on the type of motion present. For our purposes, we do not require further assumptions about the form of either $C^{v-f}(t)$ or $C^{v:f}(t)$. However, for comparison to model-free analysis, we point out that for isotropic tumbling, $C^{v:f}_{00}(t)$ is simply $\exp(-t/\tau_{global})$, and if we follow the original model-free approach, then $C^{v-f}(t)$ can be modeled as a mono-exponential function, so that eq. (18) becomes

$$C(t) = C^{v:f}(t)C^{v-f}(t) = \exp(-t/\tau_M)(S^2 + (1-S^2)\exp(-t/\tau))$$
$$= S^2 \exp(-t/\tau_M) + (1-S^2)\exp(-t/\tau_{eff})$$
$$\tau_{eff} = \frac{\tau \cdot \tau_M}{\tau + \tau_M} \approx \tau \text{ if } \tau \ll \tau_M \tag{19}$$

Note that application of model-free approach to experimental data does not actually require mono-exponentiality of the motion within frame *f*, only that for a multi-exponential correlation function, all motion is sufficiently faster than the eigenfrequencies of the spin system ($1/\tau \gg \omega$ for all $\tau$ of the internal motion and all eigenfrequencies sampled by the relaxation rate constants, $\omega$) [25,39]. In this case, fitting experimental data to the above correlation function yields $\tau_{\text{eff}}$ which is equal to the average effective correlation time of all internal motion.

*2.2.2 Separation of timescales*

In this case, we do not require isotropic tumbling, but instead assume that the two motions are separated in timescale, that is, the terms of the correlation functions for motion in *f* have equilibrated before the motion of *f* has evolved significantly. Let $t_1$ be a time at which the first motion has equilibrated but the second has not evolved significantly. Then for $t<t_1$, $\left\langle D_{p0}^2(\Omega_{\tau,t+\tau}^{\text{v:}f})\right\rangle_\tau = \delta_{0,p}$, so the total correlation function is

$$t < t_1:$$
$$C(t) = \sum_{p=-2}^{2} \underbrace{\left\langle D_{p0}^2(\Omega_{\tau,t+\tau}^{\text{v:}f})\right\rangle_\tau}_{\delta_{0,p}} \left\langle D_{0p}^2(\Omega_{\tau,t+\tau}^{\text{v}-f})\right\rangle_\tau = \left\langle D_{00}^2(\Omega_{\tau,t+\tau}^{\text{v}-f})\right\rangle_\tau . \tag{20}$$

For long correlation times ($t>t_1$), the terms $\left\langle D_{0p}^2(\Omega_{\tau,t+\tau}^{\text{v}-f})\right\rangle_\tau$ have reached their equilibrium values, which we will denote as $A_p^{\text{v}-f} = \lim_{t\to\infty}\left\langle D_{0p}^2(\Omega_{\tau,t+\tau}^{\text{v}-f})\right\rangle_\tau$, so that the correlation function becomes:

$$t > t_1:$$
$$C(t) = \sum_{p=-2}^{2} \left\langle D_{p0}^2(\Omega_{\tau,t+\tau}^{\text{v:}f})\right\rangle_\tau A_p^{\text{v}-f} . \tag{21}$$

The $A_p^{\text{v}-f}$ represent the residual tensor resulting from motion within frame *f*. Next, we define two correlation functions, $C^{\text{v:}f}(t)$ and $C^{\text{v}-f}(t)$, for which the product yields $C(t)$ at all times:

$$C^{\text{v:}f}(t) = \frac{1}{A_0^{\text{v}-f}} \sum_{p=-2}^{2} \left\langle D_{p0}^2(\Omega_{\tau,t+\tau}^{\text{v:}f})\right\rangle_\tau A_p^{\text{v}-f}$$
$$C^{\text{v}-f}(t) = \left\langle D_{00}^2(\Omega_{\tau,t+\tau}^{\text{v}-f})\right\rangle_\tau \tag{22}$$
$$C(t) = C^{\text{v:}f}(t) \cdot C^{\text{v}-f}(t)$$

For $t<t_1$, $C^{\text{v:}f}(t)=1$, so the total correlation function is simply $C^{\text{v}-f}(t)$, and for $t>t_1$, $C^{\text{v}-f}(t) = A_0^{\text{v}-f}$, which cancels with the denominator of the first term of $C^{\text{v}-f}(t)$, yielding eq.

(21), as required. Note the ratio $A_p^{v-f} / A_0^{v-f}$ retains the shape of the residual tensor but not its magnitude; the magnitude of residual tensor due to motion within *f* appears when taking the product of correlation functions (we could constructed the total correlation function as a sum where motion of *f* acts on the scaled tensor, but the time-dependent scaling of the residual tensor magnitude makes the product somewhat more robust with respect to timescale separation).

*2.2.3 Internal motion with a symmetry axis*

Our last case is a special case of the separation of timescales as discussed in the previous section. Considerable simplification can be made if we assume that the faster motion is symmetric around some axis (strictly speaking, we require minimally a 3-fold symmetry axis, such that the residual tensor of motion has no asymmetry, i.e. $\eta = 0$). We start by defining a frame for which the *z*-axis is parallel to the symmetry axis of the motion. We further define $\Omega_\tau^{v,sym}$ and $\Omega_{t+\tau}^{v,sym}$ to be the Euler angles that transform to the symmetry frame from the frame of the interaction at time $\tau$ or $t+\tau$, respectively, and $\Omega_{\tau,t+\tau}^{sym}$ to be the Euler angles rotating from the symmetry frame at time $\tau$ to its orientation at time $t+\tau$. Note that the symmetry frame replaces frame *f*, so that motion of frame *f* will no longer appear explicitly in the final set of equations. We will indicate in the following equations how terms in this approach are related to terms when no symmetry axis is present.

Then, the term $D_{0p}^2(\Omega_{\tau,t+\tau}^{v-f})$ results from motion of the vector $\mathbf{v}_Z(\tau)$ within the symmetry frame. This is achieved by expressing the motion of $\mathbf{v}_Z(\tau)$ as a rotation from the frame of $\mathbf{v}_Z(\tau)$ to the symmetry frame, and subsequently a rotation back out of the symmetry frame to the frame of $\mathbf{v}_Z(t+\tau)$. Similarly, the term $D_{p0}^2(\Omega_{\tau,t+\tau}^{v,f})$ results from motion of the vector $\mathbf{v}_Z(\tau)$ due to motion of the symmetry frame. This is achieved by rotating the vector $\mathbf{v}_Z(t+\tau)$ first back into the symmetry frame, then rotating it along with the symmetry frame itself, and finally rotating it back out of the symmetry frame. This is shown below.

$$C(t) = \sum_{p=-2}^{2} \left\langle D_{p0}^2(\Omega_{\tau,t+\tau}^{\mathbf{v},f}) D_{0p}^2(\Omega_{\tau,t+\tau}^{\mathbf{v}-f}) \right\rangle_\tau$$

$$= \sum_{p=-2}^{2} \sum_{k=-2}^{2} \sum_{j=-2}^{2} \sum_{m=-2}^{2} \left\langle \underbrace{d_{k0}^2(-\beta_{t+\tau}^{\mathbf{v},\text{sym}}) \exp(-ik\gamma_{t+\tau}^{\mathbf{v},\text{sym}}) D_{jk}^2(\Omega_{\tau,t+\tau}^{\text{sym}}) \exp(ij\gamma_{t+\tau}^{\mathbf{v},\text{sym}}) d_{pj}^2(\beta_{t+\tau}^{\mathbf{v},\text{sym}}) \exp(ip\alpha_{t+\tau}^{\mathbf{v},\text{sym}})}_{D_{p0}^2(\Omega_{\tau,t+\tau}^{\mathbf{v},f})} \right. \quad (23)$$

$$\left. \times \underbrace{\exp(-ip\alpha_{t+\tau}^{\mathbf{v},\text{sym}}) d_{mp}^2(-\beta_{t+\tau}^{\mathbf{v},\text{sym}}) \exp(-im\gamma_{t+\tau}^{\mathbf{v},\text{sym}}) \exp(im\gamma_\tau^{\mathbf{v},\text{sym}}) d_{0m}^2(\beta_\tau^{\mathbf{v},\text{sym}})}_{D_{0p}^2(\Omega_{\tau,t+\tau}^{\mathbf{v}-f})} \right\rangle$$

Rotation by $\pm\Omega_{t+\tau}^{\mathbf{v},\text{sym}}$ may easily be cancelled, and subsequently we may assume statistical independence of motion of the symmetry frame and motion within that frame.

$$C(t) = \sum_{k=-2}^{2} \sum_{m=-2}^{2} \left\langle d_{k0}^2(-\beta_{t+\tau}^{\mathbf{v},\text{sym}}) \exp(-ik\gamma_{t+\tau}^{\mathbf{v},\text{sym}}) D_{mk}^2(\Omega_{\tau,t+\tau}^{\text{sym}}) \exp(im\gamma_\tau^{\mathbf{v},\text{sym}}) d_{0m}^2(\beta_\tau^{\mathbf{v},\text{sym}}) \right\rangle_\tau$$

$$= \sum_{k=-2}^{2} \sum_{m=-2}^{2} \left\langle D_{mk}^2(\Omega_{\tau,t+\tau}^{\text{sym}}) \right\rangle_\tau \left\langle d_{k0}^2(-\beta_{t+\tau}^{\mathbf{v},\text{sym}}) \exp(-ik\gamma_{t+\tau}^{\mathbf{v},\text{sym}}) \exp(im\gamma_\tau^{\mathbf{v},\text{sym}}) d_{0m}^2(\beta_\tau^{\mathbf{v},\text{sym}}) \right\rangle_\tau$$

(24)

As before, we assume timescale separation. Suppose there is a time $t_1$ for which all times shorter than $t_1$, the symmetry frame has not evolved significantly, and at times longer than $t_1$, motion within the symmetry frame is equilibrated. Then at short times ($t < t_1$), Euler angles $\Omega_\tau^{\mathbf{v},\text{sym}}$ and $\Omega_{t+\tau}^{\mathbf{v},\text{sym}}$ are correlated (motion in the symmetry frame), but at sufficiently long times ($t > t_1$), they become uncorrelated such that

$$\left\langle d_{k0}^2(-\beta_{t+\tau}^{\mathbf{v},\text{sym}}) \exp(-ik\gamma_{t+\tau}^{\mathbf{v},\text{sym}}) \exp(im\gamma_\tau^{\mathbf{v},\text{sym}}) d_{0m}^2(\beta_\tau^{\mathbf{v},\text{sym}}) \right\rangle_\tau$$
$$= \left\langle d_{k0}^2(-\beta_{t+\tau}^{\mathbf{v},\text{sym}}) \exp(-ik\gamma_{t+\tau}^{\mathbf{v},\text{sym}}) \right\rangle_\tau \left\langle \exp(im\gamma_\tau^{\mathbf{v},\text{sym}}) d_{0m}^2(\beta_\tau^{\mathbf{v},\text{sym}}) \right\rangle_\tau$$

(25)

Then, given an $N$-fold symmetry axis, for a pair of angles $\beta$ and $\gamma$ occurring with some probability, the angles $\beta$ and $\gamma + 2\pi n/N$ must occur with the same probability for integer values of $n$. We may express this relationship of the probability densities for different angles as $P(\beta^{\mathbf{v},\text{sym}}, \gamma^{\mathbf{v},\text{sym}}) = P(\beta^{\mathbf{v},\text{sym}}, \gamma^{\mathbf{v},\text{sym}} + 2\pi n/N)$. Then, we obtain $\langle d_{k0}^2(-\beta_{t+\tau}^{\mathbf{v},\text{sym}} \exp(-ik\gamma_{t+\tau}^{\mathbf{v},\text{sym}})\rangle_\tau$ by integrating over all $\beta$ and $\gamma$ angles, multiplied by the probability of those angles, and inserting the above relationship for the probabilities.

$$\left\langle d_{k0}^2(-\beta_{t+\tau}^{v,sym})\exp(-ik\gamma_{t+\tau}^{v,sym})\right\rangle_\tau$$

$$= \int_0^\pi \int_0^{2\pi} d\beta^{v,sym} \sin\beta^{v,sym} d\gamma^{v,sym} d_{k0}^2(-\beta^{v,sym})\exp(-ik\gamma^{v,sym})P(\beta^{v,sym},\gamma^{v,sym})$$

$$= \int_0^\pi d\beta^{v,sym} \sin\beta^{v,sym} d_{k0}^2(-\beta^{v,sym}) *$$

$$\int_0^{2\pi/N} d\gamma^{v,sym} \sum_{n=0}^{N-1} \exp(-ik(\gamma^{v,sym}+2\pi n/N)) \underbrace{P(\beta^{v,sym},\gamma^{v,sym}+2\pi n/N)}_{=P(\beta^{v,sym},\gamma^{v,sym})}$$

$$= \int_0^\pi d\beta^{v,sym} \sin\beta^{v,sym} d_{k0}^2(-\beta^{v,sym}) * \qquad (26)$$

$$\int_0^{2\pi/N} d\gamma^{v,sym} P(\beta^{v,sym},\gamma^{v,sym}) \underbrace{\sum_{n=0}^{N-1} \exp(-ik(\gamma^{v,sym}+2\pi n/N))}_{=\delta_k \text{ for } N\geq 3}$$

$$= \delta_k \int_0^\pi d\beta^{v,sym} \sin\beta^{v,sym} d_{00}^2(-\beta^{v,sym}) N \underbrace{\int_0^{2\pi/N} d\gamma^{v,sym} P(\beta^{v,sym},\gamma^{v,sym})}_{=P(\beta^{v,sym})}$$

$$= \delta_k \left\langle d_{00}^2(-\beta_{t+\tau}^{v,sym})\right\rangle_\tau \qquad \text{for } N \geq 3$$

We split the inner integral into *N* steps, given by the summation. Then, we replace terms $P(\beta^{v,sym},\gamma^{v,sym}+2\pi n/N)$ with $P(\beta^{v,sym},\gamma^{v,sym})$, taking advantage of the symmetry, and move these terms out of the summation. The remaining sum of exponentials always sums to 0, except in the case that *k*=0, in which case it yields *N*. Then, if we have at least a 3-fold symmetry axis, all terms with $m \neq 0$ or $k \neq 0$ vanish for $t > t_1$, yielding

$$C(t) = \left\langle D_{00}^2(\Omega_{\tau,t+\tau}^{sym})\right\rangle_\tau \left\langle d_{00}^2(-\beta_{t+\tau}^{v,sym})\right\rangle_\tau \left\langle d_{00}^2(\beta_\tau^{v,sym})\right\rangle_\tau$$
$$= \left\langle D_{00}^2(\Omega_{\tau,t+\tau}^{sym})\right\rangle_\tau \left\langle d_{00}^2(\beta_\tau^{v,sym})\right\rangle_\tau^2 \qquad (27)$$

When $t < t_1$, $\Omega_\tau^{v,sym}$ and $\Omega_{t+\tau}^{v,sym}$ remain correlated and the symmetry axis remains fixed (frame *f* has not moved), such that $\left\langle D_{mk}^2(\Omega_{\tau,t+\tau}^{sym})\right\rangle_\tau = \delta_{m-k}$. In this case, eq. (24) becomes

$$C(t) = \sum_{m=-2}^2 \left\langle d_{m0}^2(-\beta_{t+\tau}^{v,sym})\exp(-im\gamma_{t+\tau}^{v,sym})\exp(im\gamma_\tau^{v,sym})d_{0m}^2(\beta_\tau^{v,sym})\right\rangle_\tau. \qquad (28)$$

Note that eq. (28) is equivalent to $C^{v-f}(t)$ as shown in eq. (22), only given in a different frame. Then, the appropriate definitions for $C^{v-f}(t)$ and $C^{v:f}(t)$ are

$$C^{v \cdot f}(t) = \left\langle D_{00}^2(\Omega_{\tau,t+\tau}^{\text{sym}}) \right\rangle_\tau = \left\langle d_{00}^2(\beta_{\tau,t+\tau}^{\text{sym}}) \right\rangle_\tau$$

$$C^{v-f}(t) = \left\langle D_{00}^2(\Omega_{\tau,t+\tau}^{v-f}) \right\rangle_\tau$$

$$= \sum_{m=-2}^{2} \left\langle d_{m0}^2(-\beta_{t+\tau}^{v,\text{sym}})\exp(-im\gamma_{t+\tau}^{v,\text{sym}})\exp(im\gamma_\tau^{v,\text{sym}})d_{0m}^2(\beta_\tau^{v,\text{sym}}) \right\rangle_\tau$$

(29)

Here we see that at short times ($t < t_1$), $C^{v \cdot f}(t) = 1$ and $C^{v-f}(t)$ is equal to the total correlation function found in eq. (28) (or in eq. (22)). At long correlation times, $\Omega_\tau^{v,\text{sym}}$ and $\Omega_{t+\tau}^{v,\text{sym}}$ become uncorrelated so that only the $m$=0, $k$=0 term survives in $C^{v-f}(t)$, and the product of $C^{v \cdot f}(t)$ and $C^{v-f}(t)$ is equal to the result in eq. (27). Note that for $C^{v \cdot f}(t)$, we must compute the correlation function for motion of the symmetry axis instead of computing the action of frame *f* on the original tensor, as will be discussed in section 2.4.7.

This case still requires timescale separation, as in section 2.2.2, however we no longer require a sum over multiple terms, making computation faster and simpler (we find in Appendix eq. (A20) that in addition to requiring computation of only one term vs. five, that this one term is also simpler to compute). We expect that this approach will be particularly useful for extended analyses, for example, to combine mode-analysis approaches (e.g. iRED [10,45,46]) for cross-correlating motions [10] with ROMANCE.

## 2.3  Separating more than two motions

In the previous section, we have shown that it is possible to separate two motions by defining an intermediate reference frame, *f,* either if the outer motion is isotropic tumbling (2.2.1) or if the motions are separated in timescale (2.2.2). Further simplification is possible for timescale-separated motion if an axis of symmetry (not necessarily along the bond) exists for the inner motion (2.2.3). In case we have more than two motions, these may also be separated, under the same conditions that allowed separation of two motions. We start by assuming three sets of Euler angles: $\Omega_{\tau,t+\tau}^{v-f}$, the motion of the bond within frame *f* (motion of frame *f* removed), $\Omega_{\tau,t+\tau}^{v:f-F}$, the motion of the bond due to motion of frame *f*, within frame *F* (motion of frame *F* removed), and $\Omega_{\tau,t+\tau}^{v:F}$, motion of the bond due to motion of frame *F*. Assuming these motions are statistically independent, the total correlation function is given by

$$C(t) = \sum_{p=-2}^{2}\sum_{q=-2}^{2} \left\langle D_{p0}^2(\Omega_{\tau,t+\tau}^{v:F}) \right\rangle_\tau \left\langle D_{qp}^2(\Omega_{\tau,t+\tau}^{v:f-F}) \right\rangle_\tau \left\langle D_{0q}^2(\Omega_{\tau,t+\tau}^{v-f}) \right\rangle_\tau .$$

(30)

Rather than introduce a new layer of complexity to the problem, we point out that the product of the rotations $\Omega_{\tau,t+\tau}^{v:f-F}$ and $\Omega_{\tau,t+\tau}^{v-f}$ can be collected into a single term, $\Omega_{\tau,t+\tau}^{v-F}$, so that the correlation function becomes

$$C(t) = \sum_{p=-2}^{2} \left\langle D_{p0}^{2}(\Omega_{\tau,t+\tau}^{v:F}) \right\rangle_{\tau} \left\langle D_{0p}^{2}(\Omega_{\tau,t+\tau}^{v-F}) \right\rangle_{\tau}. \tag{31}$$

One of the methods in section 2.2 then may be used to obtain the correlation function for the outer motion, $C^{v:F}(t)$:

$$C(t) = C^{v:F}(t) \cdot C^{v-F}(t)$$

$$C^{v:F}(t) = \begin{cases} \left\langle D_{00}^{2}(\Omega_{\tau,t+\tau}^{v:sym}) \right\rangle_{\tau} & \Omega_{\tau,t+\tau}^{v-F} \text{ has a symmetry axis} \\ \left\langle D_{00}^{2}(\Omega_{\tau,t+\tau}^{v:F}) \right\rangle_{\tau} & \text{Motion is symmetric about } \mathbf{v}(t) \\ \dfrac{1}{A_{p}^{v-F}} \sum_{p=-2}^{2} \left\langle D_{p0}^{2}(\Omega_{\tau,t+\tau}^{v:F}) \right\rangle_{\tau} A_{p}^{v-F} & \text{Motions are timescale separated} \end{cases} \tag{32}$$

$$C^{v-F}(t) = \left\langle D_{00}^{2}(\Omega_{\tau,t+\tau}^{v-F}) \right\rangle_{\tau} = \sum_{p=-2}^{2} \left\langle D_{p0}^{2}(\Omega_{\tau,t+\tau}^{v:f-F}) \right\rangle_{\tau} \left\langle D_{0p}^{2}(\Omega_{\tau,t+\tau}^{v-f}) \right\rangle_{\tau}$$

Regardless of the method used to separate the correlation function for the outer motion ($C^{v:F}(t)$), we always have the same functional form for the remaining inner motion ($C^{v-F}(t)$, as given in eq. (32)). As we showed in sections 2.2.1-2.2.3, this equation may then be further separated into two more motions, only now we must apply the analysis within a frame where motion of $F$ is removed, such that

$$C(t) = C^{v:F}(t) \cdot C^{v-F}(t) = C^{v:F}(t) \cdot C^{v:f-F}(t) \cdot C^{v-f}(t). \tag{33}$$

This principle applies to an arbitrary number of motions, where we always separate out the outer motion first, and initially treat the product of all inner motions using a single set of Euler angles. As before, at each step we require statistical independence of the outer motion and net influence of the inner set of motions, and we must meet one of the requirements set out in sections 2.2.1-2.2.3.

## 2.4 Extracting Euler Angles for correlation functions

We have established that it is possible to separate the total correlation function, $C(t)$, into a product of terms resulting from two or more motions, under the condition of statistical independence, and either timescale separation or symmetry around $\mathbf{v}_z(\tau)$ (i.e. isotropic tumbling). Then, to obtain the correlation functions, we must define reference frames and then determine how to extract the required Euler angles (see Table 2) from the coordinates of the various frames.

*2.4.1 Defining frames 1: Frame of the NMR interaction*

In this study, we utilize frames defined by the interaction tensor (CSA, dipole, etc.), and frames used to separate individual contributions to the total motion, which are defined by positions of atoms in the MD simulation. For example, the time-dependent direction of the interaction tensor, $\mathbf{v}_z(\tau)$, is required to determine the total correlation function. This correlation function, $C(t)$, is given in eq. (10), where one must determine the angles $\beta^{\mathbf{v}}_{\tau,t+\tau}$, which rotate $\mathbf{v}^{\mathbf{v}}_z(\tau)$ (that is, $\mathbf{v}_z(\tau)$ in its own frame, such that $\mathbf{v}^{\mathbf{v}}_z(\tau) = [0,0,1]'$) to $\mathbf{v}^{\mathbf{v}}_z(t+\tau)$, which is the interaction at time $t+\tau$ given in the frame of the interaction at time $\tau$.

First, we note that the direction of the principle axis of the interaction tensor only defines the z-axis of its frame, whereas the x- and y-axes are not defined. For calculating $C(t)$, we only require the angle $\beta^{\mathbf{v}}_{\tau,t+\tau}$, which is the angle between the z-axes of the frames of the interaction at times $\tau$ and $t+\tau$, and so we do not require a definition for x and y. On the other hand, when separating motion into contributions based strictly on timescale separation (section 2.2.2), we note that terms appearing in eq. (22) depend on terms $A^{\mathbf{v}-f}_p$, which depend on $\gamma^{\mathbf{v}-f}_{\tau,t+\tau}$, and $\left\langle D^2_{p0}(\Omega^{\mathbf{v}:f}_{\tau,t+\tau})\right\rangle_\tau$, which depend on $\alpha^{\mathbf{v}:f}_{\tau,t+\tau}$. These terms appear in the correlation functions for separated motion because the relative directions of the motion of frame *f* and of the residual tensor due to motion in *f* ($A^{\mathbf{v}-f}_p$) are important for determining the influence of the motion of frame *f* on the total correlation function. For example, suppose motion within frame *f* rotates freely around some axis. Then, the residual tensor will point along that axis. If the motion of the frame rotates around the same axis, then it will have no influence on the total correlation function, but if it were to rotate perpendicular to the original motion, then it would result in further decay of the correlation function. Therefore, to properly account for the combined effects of multiple motions, we need to ensure that the x- and y-axes remain well defined.

In order to fully define the frame of the interaction, we use the following approach: first, we define the z-axis; for relaxation via dipole couplings this will typically be a bond vector. As before, this vector is denoted $\mathbf{v}_z(\tau)$, and for a dipole coupling, can be calculated simply from the difference in positions of the dipole-coupled nuclei, followed by normalization of the length. The directions of the x- and y-axes are arbitrary, but must be nonetheless consistently defined so that motion resulting from motion of frame *f* can be correctly applied to the residual tensors resulting from motion within frame *f*. Then, if we want to calculate the correlation function for the motion of a one bond H–X dipole tensor, we could take $\mathbf{v}_Z(\tau)$ to point in the direction of the H–X bond, and we let the bond between X and some third atom, Y, define the positive *xz*-plane. The resulting vector will be referred to as $\mathbf{v}_{XZ}(\tau)$. We will show below that vectors $\mathbf{v}_X(\tau)$ and $\mathbf{v}_Y(\tau)$ can be easily obtained from $\mathbf{v}_Z(\tau)$ and $\mathbf{v}_{XZ}(\tau)$, but first we discuss defining frames *f*.

*2.4.2 Defining frames 2: Frames to separate motion*

Once we have the frame of the interaction defined, via $\mathbf{v}_Z(\tau)$ and $\mathbf{v}_{XZ}(\tau)$, we next need to define frames (*f*) that may be used to separate individual contributions to the overall motion. As with the interaction frame, we will use one vector to define the *z*-axis of frame *f*, and a second vector to define the positive *xz*-plane. We will refer to these as $v_Z^f(\tau)$ and $v_{XZ}^f(\tau)$. Note that, unlike the interaction frame, it will not always be necessary to define the *xz*-plane. For example, if one considers relaxation of a carbon in the acyl chain of a lipid, then one might want to separate contributions of overall motion of that chain, defining $v_Z^f(\tau)$ to lie along the largest of the principle components of the moment of inertia tensor [38]. While $v_{XZ}^f(\tau)$ could be defined as the middle or smallest component, the narrowness of the chain may cause these components to swap suddenly, yielding an unstable definition for $v_{XZ}^f(\tau)$. To avoid such instability, the appropriate internal frame (*f*) would use the *z*-axis of the MOI to define $v_Z^f(\tau)$ and would not define $v_{XZ}^f(\tau)$. On the other hand, one might try to separate overall motion of an α-helix in a protein from motion within the helix. In this case, one could choose several reference atoms within the helix and determine the rotation matrix that aligns those atoms between different frames of the trajectory. Then, one can extract both $v_Z^f(\tau)$ and $v_{XZ}^f(\tau)$ from the rotation matrix (in fact, the columns of the rotation matrix yield $v_X^f(\tau)$, $v_Y^f(\tau)$, and $v_Z^f(\tau)$).

Note that only the relative directions between these vectors at different times (e.g. $\tau$, $t+\tau$) will influence the results, yielding a good deal of flexibility when defining the vectors for a frame *f*. However, $v_Z(\tau)$ and $v_{XZ}(\tau)$ should first, of course, capture the dynamics of the motion of interest, they should never become equal (which would prevent defining the *x*- and *y*-axes), and they should not exhibit discontinuities (for example, attempting to align an unstructured region of a protein might lead to sudden, sharp changes in the rotation matrix for alignment, since alignment to an unstructured region might not be well defined).

*2.4.3 Obtaining Euler angles 1: Lab frame*

Once we have the interaction frame and various reference frames, *f*, defined and obtain the time dependent vectors, e.g. $\mathbf{v}_Z(\tau)$, $\mathbf{v}_{XZ}(\tau)$, $v_Z^f(\tau)$, $v_{XZ}^f(\tau)$, we need to extract a number of different sets of Euler angles. The first set of Euler angles consists of those angles required to rotate from the lab frame (where the vectors are initially expressed), into the frame of the interaction or into the reference frame, *f*. For example, given $v_Z^f(\tau)$ and $v_{XZ}^f(\tau)$, which define the orientation of frame *f* as a function of time, the corresponding Euler angles, $\Omega_\tau^f = \{\alpha_\tau^f, \beta_\tau^f, \gamma_\tau^f\}$ can be calculated according to

$$\begin{aligned}
v_Z^f(\tau) &= [x_Z^f(\tau), y_Z^f(\tau), z_Z^f(\tau)] \\
\cos\gamma_\tau^f &= x_Z^f(\tau) / \sqrt{(x_Z^f(\tau))^2 + (y_Z^f(\tau))^2} \\
\sin\gamma_\tau^f &= y_Z^f(\tau) / \sqrt{(x_Z^f(\tau))^2 + (y_Z^f(\tau))^2} \\
\cos\beta_\tau^f &= z_Z^f(\tau) \\
\sin\beta_\tau^f &= \sqrt{1 - (z_Z^f(\tau))^2}
\end{aligned} \qquad (34)$$

Note that we have assumed $v_Z^f(\tau)$ to be normalized. Then, $v_Z^f(\tau)$ defines the angles $\beta_\tau^v$ and $\gamma_\tau^v$, whereas $\alpha_\tau^v$ also depends on $v_{XZ}^f(\tau)$. Thus we first rotate $v_{XZ}^f(\tau)$ by $\beta_\tau^v$ and $\gamma_\tau^v$, and then extract $\alpha_\tau^f$ from the result.

$$\begin{aligned}
v_{XZ}^{f*}(\tau) &= [x_{XZ}^{f*}(\tau), y_{XZ}^{f*}(\tau), z_{XZ}^{f*}(\tau)] = \mathbf{R}_Y(-\beta_\tau^f) \cdot \mathbf{R}_Z(-\gamma_\tau^f) \cdot v_{XZ}^f(\tau) \\
\cos\alpha_\tau^f &= x_{XZ}^{f*}(\tau) / \sqrt{(x_{XZ}^{f*}(\tau))^2 + (y_{XZ}^{f*}(\tau))^2} \\
\sin\alpha_\tau^f &= y_{XZ}^{f*}(\tau) / \sqrt{(x_{XZ}^{f*}(\tau))^2 + (y_{XZ}^{f*}(\tau))^2}
\end{aligned} \qquad (35)$$

If $v_{XZ}^f(\tau)$ is not defined for a frame, then we set $\alpha_\tau^f = -\gamma_\tau^f$. The following relationships are satisfied by the Euler angles:

$$\begin{aligned}
v_Z^f(\tau) &= \mathbf{R}_Z(-\alpha_\tau^f) \cdot \mathbf{R}_Y(-\beta_\tau^f) \cdot \mathbf{R}_Z(-\gamma_\tau^f) \cdot v_Z^f(\tau) = [0,0,1]' \\
v_{XZ}^f(\tau) &= \mathbf{R}_Z(-\alpha_\tau^f) \cdot \mathbf{R}_Y(-\beta_\tau^f) \cdot \mathbf{R}_Z(-\gamma_\tau^f) \cdot v_{XZ}^f(\tau) = [\sqrt{1-(z_{XZ}^f)^2}, 0, z_{XZ}^f]'
\end{aligned} \qquad (36)$$

In frame *f* at time $\tau$, $v_Z^f(\tau)$ lies along the *z*-axis, and $v_{XZ}^f(\tau)$ lies in the *xz*-plane. Note that we apply a passive rotation to bring the vectors from the lab frame into frame *f*. The angles that bring vectors $\mathbf{v}_Z(\tau)$ and $\mathbf{v}_{XZ}(\tau)$ from their own frame into the lab frame are obtained similarly, and are denoted as $\Omega_\tau^v = \{\alpha_\tau^v, \beta_\tau^v, \gamma_\tau^v\}$.

The Euler angles themselves may be obtained with a two-argument arctangent (to ensure angles are returned in the correct quadrant). However, we only actually need the sines and cosines of the angles for further calculation, so we leave them in the form above. Once we have the Euler angles, it is straightforward to also obtain vectors $v_X(\tau)$ and $v_Y(\tau)$:

$$v_X(\tau) = \mathbf{R}_{ZYZ}(\Omega_\tau^f) \cdot [1,0,0]'$$
$$v_Y(\tau) = \mathbf{R}_{ZYZ}(\Omega_\tau^f) \cdot [0,1,0]'. \quad (37)$$
$$v_Z(\tau) = \mathbf{R}_{ZYZ}(\Omega_\tau^f) \cdot [0,0,1]'$$

$v_X(\tau)$, $v_Y(\tau)$, and $v_Z(\tau)$ are just the respective columns of $\mathbf{R}_{ZYZ}(\Omega_\tau^v)$.

### 2.4.4 Removing motion of frame f or F

One of the first procedures we will need to perform is the removal of motion of a given frame, in order to characterize motion within that frame. This is achieved by obtaining the sines and cosines of the Euler angles for frame *f* or *F* ($\Omega_\tau^F$, eqs. (34)-(35)), and applying the inverse rotation matrix to either vectors in the lab frame representing some inner frame, *f* ($v_Z^f(\tau)$, $v_{XZ}^f(\tau)$), or representing the frame of the interaction itself ($\mathbf{v}_Z(\tau)$, $\mathbf{v}_{XZ}(\tau)$). In some of the subsequent equations, we replace subscripts X and XZ with $\zeta$, to indicate that the calculation may be applied to either vector.

$$\mathbf{v}_\zeta^{-F}(\tau) = \mathbf{R}_{ZYZ}^{-1}(\Omega_\tau^F) \cdot \mathbf{v}_\zeta(\tau) = \mathbf{R}_Z(-\alpha_\tau^F) \cdot \mathbf{R}_Y(-\beta_\tau^F) \cdot \mathbf{R}_Z(-\gamma_\tau^F) \cdot \mathbf{v}_\zeta(\tau)$$
$$v_\zeta^{f-F} = \mathbf{R}_{ZYZ}^{-1}(\Omega_\tau^F) \cdot v_\zeta^f(\tau) = \mathbf{R}_Z(-\alpha_\tau^F) \cdot \mathbf{R}_Y(-\beta_\tau^F) \cdot \mathbf{R}_Z(-\gamma_\tau^F) \cdot v_\zeta^f(\tau)$$ . (38)

The first formula yields the motion of the interaction, now in the frame *F*. Since this removes the influence of the motion of *F*, we denote this with the superscript –F. Similarly, the motion of frame *f*, now in the frame of *F* is denoted with a superscript *f–F*, that is motion of *f* with motion of *F* removed. This procedure is illustrated in Fig. 4A-B.

### 2.4.5 Obtaining Euler angles 2: Motion within frame f

In order to separate the total correlation function into contributions from two or more motions, we need to obtain the correlation function for motion within frame *f*, given by $\langle D_{00}^2(\Omega_{\tau,t+\tau}^{v-f}) \rangle_\tau$ and we need the residual tensor for motion with frame *f*, given by the $A_p^{v-f} = \lim_{t \to \infty} \langle D_{0p}^2(\Omega_{\tau,t+\tau}^{v-f}) \rangle_\tau$ (eq. (22)). To find these terms, we calculate the Euler angles $\Omega_{\tau,t+\tau}^{v-f} = \{\alpha_{\tau,t+\tau}^{v-f}, \beta_{\tau,t+\tau}^{v-f}, \gamma_{\tau,t+\tau}^{v-f}\}$, and insert them into elements of the Wigner rotation matrix (Table 2). To obtain the angles, we first need to remove the motion of frame *f*, as done in eq. (38), to obtain $\mathbf{v}_Z^{v-f}(\tau)$, $\mathbf{v}_{XZ}^{v-f}(\tau)$. Then, we note the following relationships: first, if a vector initially along the *z*-axis of some frame is rotated by Euler angles $\Omega = \{\alpha,\beta,\gamma\}$, the resulting vector is

$$\mathbf{R}_{ZYZ}(\Omega) \cdot \begin{pmatrix} 0 \\ 0 \\ 1 \end{pmatrix} = \begin{pmatrix} \sin\beta\cos\gamma \\ \sin\beta\sin\gamma \\ \cos\beta \end{pmatrix}. \qquad (39)$$

Then, to extract the $\beta^{v-f}_{\tau,t+\tau}$ and $\gamma^{v-f}_{\tau,t+\tau}$ angles, we just need to find $\mathbf{v}^{v-f}_Z(t+\tau)$, that is, the direction of the interaction at time $t+\tau$, given in the frame of the interaction at time $\tau$, since $\mathbf{v}^{v-f}_Z(t+\tau) = \mathbf{R}_{ZYZ}(\Omega^{v-f}_{\tau,t+\tau}) \cdot [0,0,1]$ (eq. (11)). A simple procedure to obtain $\mathbf{v}^{v-f}_Z(t+\tau)$ without requiring the calculation of Euler angles $\Omega^{v-f}_\tau$ at all times, $\tau$, is to project $\mathbf{v}^{v-f}_Z(t+\tau)$ into the frame of the interaction at time $\tau$. Since the vectors $\mathbf{v}^{-f}_X(\tau)$, $\mathbf{v}^{-f}_Y(\tau)$, and $\mathbf{v}^{-f}_Z(\tau)$ define the frame of the interaction at time $\tau$ within frame $f$ (motion of $f$ removed), the projection is performed by calculating.

$$\mathbf{v}^{v-f}_Z(t+\tau) = \begin{pmatrix} \mathbf{v}^{-f}_X(\tau) \\ \mathbf{v}^{-f}_Y(\tau) \\ \mathbf{v}^{-f}_Z(\tau) \end{pmatrix} \cdot \mathbf{v}^{-f}_Z(t+\tau) = \mathbf{R}_{ZYZ}(\Omega^{v-f}_{\tau,t+\tau}) \cdot \begin{pmatrix} 0 \\ 0 \\ 1 \end{pmatrix} = \begin{pmatrix} \sin\beta^{v-f}_{\tau,t+\tau}\cos\gamma^{v-f}_{\tau,t+\tau} \\ \sin\beta^{v-f}_{\tau,t+\tau}\sin\gamma^{v-f}_{\tau,t+\tau} \\ \cos\beta^{v-f}_{\tau,t+\tau} \end{pmatrix}. \qquad (40)$$

For completeness, we point out that the angle $\alpha^{v-f}_{\tau,t+\tau}$ may be obtained in a similar manner (although it will not appear in the terms $\left\langle D^2_{0p}(\Omega^{v-f}_{\tau,t+\tau}) \right\rangle_\tau$). To obtain $\alpha^{v-f}_{\tau,t+\tau}$, we invert the projection to get the time-reversed angles, $\beta^{v-f}_{t+\tau,\tau}$ and $\gamma^{v-f}_{t+\tau,\tau}$.

$$\begin{pmatrix} \mathbf{v}^{-f}_X(t+\tau) \\ \mathbf{v}^{-f}_Y(t+\tau) \\ \mathbf{v}^{-f}_Z(t+\tau) \end{pmatrix} \cdot \mathbf{v}^{-f}_Z(\tau) = \begin{pmatrix} \sin\beta^{v-f}_{t+\tau,\tau}\cos\gamma^{v-f}_{t+\tau,\tau} \\ \sin\beta^{v-f}_{t+\tau,\tau}\sin\gamma^{v-f}_{t+\tau,\tau} \\ \cos\beta^{v-f}_{t+\tau,\tau} \end{pmatrix} = \begin{pmatrix} -\cos\alpha^{v-f}_{\tau,t+\tau}\sin\beta^{v-f}_{\tau,t+\tau} \\ \sin\alpha^{v-f}_{\tau,t+\tau}\sin\beta^{v-f}_{\tau,t+\tau} \\ \cos\beta^{v-f}_{\tau,t+\tau} \end{pmatrix}. \qquad (41)$$

The last relationship results from the fact that we can invert the rotation by changing the order and sign of the Euler angles ($\alpha^{v-f}_{\tau,t+\tau} = -\gamma^{v-f}_{t+\tau,\tau}$, $\beta^{v-f}_{\tau,t+\tau} = -\beta^{v-f}_{t+\tau,\tau}$, $\gamma^{v-f}_{\tau,t+\tau} = -\alpha^{v-f}_{t+\tau,\tau}$). Although we can extract the angles themselves from the above equations, we may leave them in this form to calculate elements of $\left\langle D^2_{0p}(\Omega^{v-f}_{\tau,t+\tau}) \right\rangle_\tau$, as seen in Table 2.

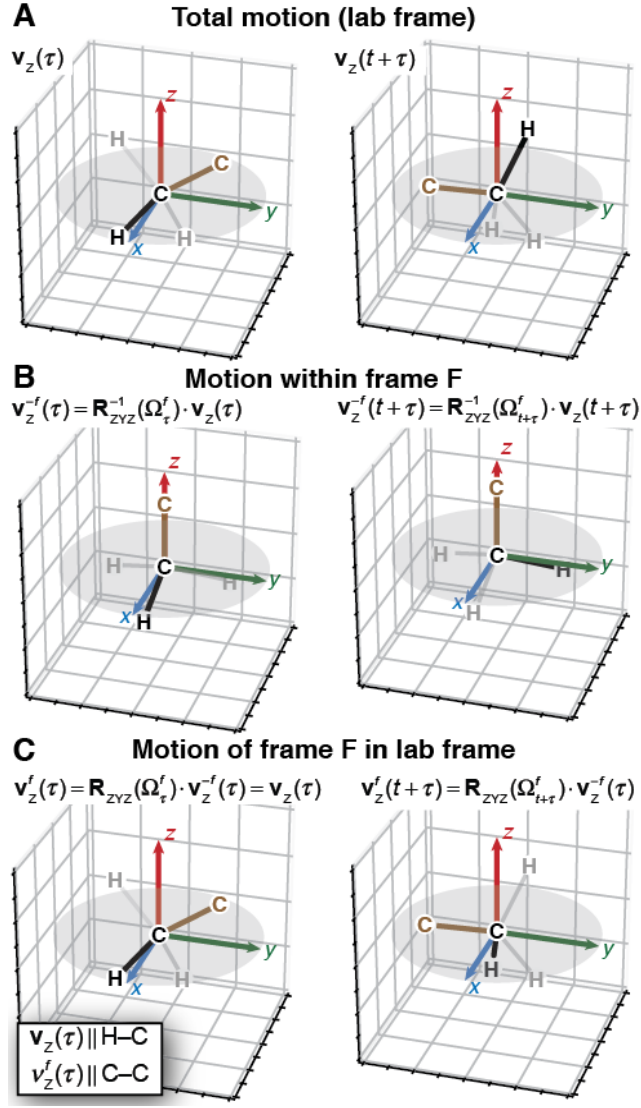

**Fig. 4.** Separating motion with frames. We take a methyl group as an example, where we are interested in the motion of one of the H–C bonds (black, bold); this bond defines $\mathbf{v}_z(\tau)$. We separate methyl rotation from overall motion of the methyl group, defined by motion of the C–C bond (brown). Therefore, we define the frame *f* with the C–C bond ($v_z^f(\tau)\|$ C–C). **A** shows the orientation of the methyl group in the lab frame at time $\tau$ (left) and at $t+\tau$ (right). **B** shows the motion of the H–C bond within the frame *f*, obtained by rotating the methyl group such that the C–C bond ($v_z^f(\tau)$) lies along the z-axis (eq. (38)) at both times $\tau$ and $t+\tau$. **C** shows the motion of the H–C bond due to the motion of frame *f*. At time $\tau$, the orientation is the same as in the lab frame but at time $t+\tau$, we obtain reorientation of the H–C bond due to motion of frame *f* between times $\tau$ and $t+\tau$ (eq. (43)).

### 2.4.6 Obtaining Euler angles 3: Motion of frame f

To obtain the correlation function resulting from motion of frame *f*, we need to obtain the corresponding Euler angles $\Omega_{\tau,t+\tau}^{v:f}$, which give the rotation of the tensor from time $\tau$ to time $t+\tau$ due to motion of frame *f* in the frame of the tensor at time $\tau$, in order to calculate terms $\left\langle D_{p0}^2(\Omega_{\tau,t+\tau}^{v:f})\right\rangle_\tau$. For cases that we have two or more frames, we will need to calculate these Euler angles after first removing motion of another frame, *F*, where these angles are denoted $\Omega_{\tau,t+\tau}^{v:f-F}$. We start with vectors $\mathbf{v}_\zeta^{-F}(\tau)$ and $v_\zeta^{f-F}(\tau)$, which describe the motion of the interaction in frame *F* and the motion of frame *f* in frame *F*, respectively (for removal of motion of *F*, see eq. (38)). While we can obtain Euler angles describing the motion of frame *f* in frame *F* ($\Omega_\tau^{f-F}$, see eqs. (34), (35)), these are not the correct angles since we require motion of the frame *f* where motion of *F* is removed, but represented in the frame of the interaction at time $\tau$.

We work backwards from our goal: Suppose we already have vectors that move only due to motion of frame *f* and have motion of frame *F* removed, and furthermore are given in the frame of the interaction at time $\tau$. We denote these vectors as $\mathbf{v}_\zeta^{v:f-F}(t+\tau)$ (the superscript indicates being in the frame of the interaction at time $\tau$ where motion is due to frame *f*, while motion of frame *F* is removed). Note that the vector $\mathbf{v}_Z^{v:f-F}(\tau)$ is expressed in its own frame, so always equals [0,0,1]'. It follows that the Euler angles $\beta_{\tau,t+\tau}^{v:f-F}$ and $\gamma_{\tau,t+\tau}^{v:f-F}$ can be extracted from $\mathbf{v}_Z^{v:f-F}(t+\tau)$ as in eq. (39) ($\alpha_{\tau,t+\tau}^{v:f-F}$ can be similarly obtained, see below). However, we first need to obtain $\mathbf{v}_Z^{v:f-F}(t+\tau)$. We can do so if we have vectors $\mathbf{v}_{XZ}^{f-F}(t+\tau)$, and $\mathbf{v}_Z^{f-F}(t+\tau)$ which give the direction of the interaction in the frame *F*, where the rotation of the interaction between time $\tau$ and $t+\tau$ is due only to the motion of frame *f* within *F*. Note that the difference between $\mathbf{v}_Z^{f-F}(t+\tau)$ and $\mathbf{v}_Z^{v:f-F}(t+\tau)$ is that the former is defined in frame *F* at time $t+\tau$, and the latter is defined in the frame of the interaction at time $\tau$ (in the frame of $\mathbf{v}_Z^{v:f-F}(\tau)$). Furthermore, we should differentiate $\mathbf{v}_Z^{f-F}(\tau)$ from $\mathbf{v}_Z^{-F}(\tau)$, where the former only moves due to motion of *f*, but the latter moves both due to motion in *f* and motion of *f* (both have motion of *F* removed). Then, we can in principle obtain $\mathbf{v}_Z^{v:f-F}(t+\tau)$ by projecting $\mathbf{v}_Z^{f-F}(t+\tau)$ into the frame of the interaction at time $\tau$, defined by $\mathbf{v}_\zeta^{f-F}(\tau)$ (compare this approach to eq. (40)).

$$\mathbf{v}_Z^{v:f-F}(t+\tau) = \begin{pmatrix} \mathbf{v}_X^{f-F}(\tau) \\ \mathbf{v}_Y^{f-F}(\tau) \\ \mathbf{v}_Z^{f-F}(\tau) \end{pmatrix} \cdot \mathbf{v}_Z^{f-F}(t+\tau) = \mathbf{R}_{ZYZ}(\Omega_{\tau,t+\tau}^{v:f-F}) \cdot \begin{pmatrix} 0 \\ 0 \\ 1 \end{pmatrix} = \begin{pmatrix} \sin\beta_{\tau,t+\tau}^{v:f-F} \cos\gamma_{\tau,t+\tau}^{v:f-F} \\ \sin\beta_{\tau,t+\tau}^{v:f-F} \sin\gamma_{\tau,t+\tau}^{v:f-F} \\ \cos\beta_{\tau,t+\tau}^{v:f-F} \end{pmatrix}. \quad (42)$$

As one sees, supposing we have $\mathbf{v}_Z^{f-F}(\tau)$, we may also obtain the desired Euler angles (time reversal as in eq. (41) can also provide $\alpha_{\tau,t+\tau}^{v:f-F}$).

We still must obtain $\mathbf{v}_Z^{f-F}(\tau)$ and $\mathbf{v}_Z^{f-F}(t+\tau)$. Note that $\mathbf{v}_Z^{f-F}(t+\tau)$ is the result of rotation of frame *f* between times $\tau$ and $t+\tau$, therefore we may take $\mathbf{v}_Z^{-F}(\tau)$, and rotate it first from frame *F* into the frame of *f*, by applying $\mathbf{R}_{ZYZ}^{-1}(\Omega_\tau^{f-F})$, and subsequently rotating back into frame *F* from frame *f*, by applying $\mathbf{R}_{ZYZ}(\Omega_{t+\tau}^{f-F})$, thus obtaining the net rotation due to motion of frame *f*. The angles $\Omega_\tau^{f-F}$ are obtained from $v_Z^{f-F}(\tau)$ and $v_{XZ}^{f-F}(\tau)$ via eqs. (34), (35).

$$\mathbf{v}_Z^{f-F}(t+\tau) = \mathbf{R}_{ZYZ}(\Omega_{t+\tau}^{f-F}) \cdot \underbrace{\mathbf{R}_{ZYZ}^{-1}(\Omega_\tau^{f-F}) \cdot \mathbf{v}_Z^{-F}(\tau)}_{\mathbf{v}_Z^f(\tau)} \tag{43}$$

This equation indicates that $\mathbf{v}_Z^{f-F}(\tau)$ is equal to $\mathbf{v}_Z^{-F}(\tau)$, since replacing $t+\tau$ in the above equation with $\tau$ causes the two rotation matrices to cancel. These rotations are illustrated in Fig. 4C. It may be useful for understanding to compare the expression for $\mathbf{v}_Z^{f-F}(t+\tau)$ to the formula for $\mathbf{v}_Z^{-F}(t+\tau)$, where the latter reorients both due to motion in frame *f* and of frame *f*.

$$\mathbf{v}_Z^{-F}(t+\tau) = \mathbf{R}_{ZYZ}(\Omega_{t+\tau}^{f-F}) \cdot \underbrace{\left[ \mathbf{R}_{ZYZ}(\Omega_{t+\tau}^{v-f}) \cdot \mathbf{R}_{ZYZ}^{-1}(\Omega_\tau^{v-f}) \right] \cdot \underbrace{\mathbf{R}_{ZYZ}^{-1}(\Omega_\tau^{f-F}) \cdot \mathbf{v}_Z^{-F}(\tau)}_{\mathbf{v}_Z^f(\tau)}}_{\mathbf{v}_Z^f(t+\tau)} \tag{44}$$

$$=[0,0,1]'$$

We are guided through the steps by the braces underneath the equation. We start with the interaction in frame *F* at time $\tau$ (motion of frame *F* is removed in all steps here, where we indicate its removal in the superscripts only for generality). To arrive at the orientation in frame *F* at time $t+\tau$, we first remove motion of frame *f* at time $\tau$, followed by removal of motion of the tensor in frame *f* (yielding a vector pointing along *z*). Then, we add motion in frame *f* at time $t+\tau$, and finally add motion of frame *f* at time $t+\tau$. By comparing eqs. (43) and (44), we see that the results are similar, but terms in square brackets in eq. (44) are missing from eq. (43). The missing term, $\mathbf{R}_{ZYZ}(\Omega_{t+\tau}^{v-f}) \cdot \mathbf{R}_{ZYZ}^{-1}(\Omega_\tau^{v-f})$, is just the rotation of the vector due to motion in frame *f* between times $\tau$ and $t+\tau$, so that we can see that we have successfully removed this motion, and only motion of frame *f* remains.

Finally, we may use eq. (42) and project $\mathbf{v}_Z^{f-F}(t+\tau)$ into the frame of $\mathbf{v}_Z^{f-F}(\tau)$ (equal to $\mathbf{v}_Z^{-F}(\tau)$), while noting that $\mathbf{R}_{ZYZ}^{-1}(\Omega_\tau^{f-F}) \cdot \mathbf{v}_Z^{-F}(\tau) = \mathbf{v}_Z^{-f}(\tau)$. Doing so, we obtain the desired Euler angles:

$$\begin{pmatrix} \mathbf{v}_X^{-F}(\tau) \\ \mathbf{v}_Y^{-F}(\tau) \\ \mathbf{v}_Z^{-F}(\tau) \end{pmatrix} \cdot \mathbf{R}_{ZYZ}(\Omega_{t+\tau}^{f-F}) \cdot \mathbf{v}_Z^{-f}(\tau) = \begin{pmatrix} \sin\beta_{\tau,t+\tau}^{v:f-F} \cos\gamma_{\tau,t+\tau}^{v:f-F} \\ \sin\beta_{\tau,t+\tau}^{v:f-F} \sin\gamma_{\tau,t+\tau}^{v:f-F} \\ \cos\beta_{\tau,t+\tau}^{v:f-F} \end{pmatrix}. \quad (45)$$

As in eq. (41), we may obtain $\alpha_{\tau,t+\tau}^{v:f-F}$ by inverting the direction of the projection (we require $\alpha_{\tau,t+\tau}^{v:f-F}$ to calculate $\langle D_{p0}^2(\Omega_{\tau,t+\tau}^{v:f-F}) \rangle_\tau$).

$$\begin{pmatrix} \mathbf{v}_X^{-F}(t+\tau) \\ \mathbf{v}_Y^{-F}(t+\tau) \\ \mathbf{v}_Z^{-F}(t+\tau) \end{pmatrix} \cdot \mathbf{R}_{ZYZ}(\Omega_\tau^{f-F}) \cdot \mathbf{v}_Z^{-f}(t+\tau) = \begin{pmatrix} \sin\beta_{t+\tau,\tau}^{v:f-F} \cos\gamma_{t+\tau,\tau}^{v:f-F} \\ \sin\beta_{t+\tau,\tau}^{v:f-F} \sin\gamma_{t+\tau,\tau}^{v:f-F} \\ \cos\beta_{t+\tau,\tau}^{v:f-F} \end{pmatrix} = \begin{pmatrix} -\cos\alpha_{\tau,t+\tau}^{v:f-F} \sin\beta_{\tau,t+\tau}^{v:f-F} \\ \sin\alpha_{\tau,t+\tau}^{v:f-F} \sin\beta_{\tau,t+\tau}^{v:f-F} \\ \cos\beta_{\tau,t+\tau}^{v:f-F} \end{pmatrix}. \quad (46)$$

Note that in this section, for generality, we have assumed that some outer motion or motions may be removed by defining an additional outer frame $F$. However, if this is the outermost motion, then no frame $F$ will be defined and may be simply replaced by the static lab frame, i.e. $\Omega_\tau^{f-F} = \Omega_\tau^f$, $\mathbf{v}_\zeta^{-F}(\tau) = \mathbf{v}_\zeta(\tau)$, etc.

*2.4.7 Motion due to frame f, when motion in frame f is symmetric*

As was noted in section 2.2.3, if internal motion within some frame $f$ is symmetric about a given axis (not necessarily the original direction of the interaction), then the correlation for the motion of frame $f$, $C^{v:f-F}(t)$, may be obtained from the motion of the symmetry axis. This simplifies the calculation of the correlation function in a number of ways, as we will see below.

In this case, we must calculate $\langle d_{00}^2(\beta_{\tau,t+\tau}^{sym}) \rangle_\tau$ (note, we may also first remove motion of another frame $F$, then requiring $\langle d_{00}^2(\beta_{\tau,t+\tau}^{sym-F}) \rangle_\tau$). This only requires the angle $\beta_{\tau,t+\tau}^{sym}$, so that we need to obtain the z-axis of the symmetry frame as a function of time, denoted $\mathbf{v}_Z^{sym}(\tau)$. Note that $\mathbf{v}_Z^{sym}(\tau)$ reorients as the result of the motion of frame $f$, although the two frames may not have the same z-axis. Then, to obtain $\mathbf{v}_Z^{sym}(\tau)$ we first calculate the residual interaction tensor within frame $f$. The components of this tensor are given by

$$A_p^{-f} = \langle D_{0p}^2(\Omega_\tau^{v-f}) \rangle_\tau. \quad (47)$$

$\Omega_\tau^{v-f}$ are simply the Euler angles describing the direction of the tensor after removing motion of frame $f$ (apply eq. (34) to $\mathbf{v}_Z^{-f}(\tau)$). Note these terms are defined in frame $f$, not in the frame of the tensor at time $\tau$ (compare to $A_p^{v-f}$, eq. (21)). Then, we must extract the z-direction of this averaged tensor, which we denote $\mathbf{v}_Z^{\text{resid}-f}$ (see Appendix 1 for obtaining $\mathbf{v}_Z^{\text{resid}-f}$). This vector is time independent in frame $f$. $\mathbf{v}_Z^{\text{sym}}(\tau)$ may be obtained simply by applying the motion of frame $f$ to this residual tensor, in other words, by rotating $\mathbf{v}_Z^{\text{resid}-f}$ with $\Omega_\tau^{-f}$:

$$\mathbf{v}_Z^{\text{sym}}(\tau) = \mathbf{R}_Z(\gamma_\tau^{f-F}) \cdot \mathbf{R}_Z(\beta_\tau^{f-F}) \cdot \mathbf{R}_Z(\alpha_\tau^{f-F}) \cdot \mathbf{v}_Z^{\text{resid}-f}. \tag{48}$$

The Euler angles for this frame (only $\beta_{\tau,t+\tau}^{\text{sym}}$ is required) may be obtained as in eq. (40), yielding

$$\mathbf{v}_Z^{\text{sym}}(\tau) \cdot \mathbf{v}_Z^{\text{sym}}(t+\tau) = \cos\beta_{\tau,t+\tau}^{\text{sym}}. \tag{49}$$

Once $\cos\beta_{\tau,t+\tau}^{\text{sym}}$ is obtained, we can already calculate the required correlation function,

$$C^{v-f}(t) = \left\langle D_{00}^2(\Omega_{\tau,t+\tau}^{\text{sym}}) \right\rangle_\tau.$$

## 2.5  Summary of calculation of correlation functions

At this stage, we know in principle how to calculate all correlation functions and residual tensors required for separating the total motion into components. We summarize the procedures here, where the exact steps depend on whether a motion is the innermost motion ($C^{v-f}(t)$, that is, motion of the interaction tensor with motion of frame $f$ removed), an intermediate motion ($C^{v:f-F}(t)$, motion of the interaction tensor due to motion of frame $f$, with motion of frame $F$ removed), and the outermost motion ($C^{v:F}(t)$, motion of the interaction tensor due to motion of frame $F$, with no other motion removed). Note that multiple intermediate frames may be used, where these all follow the same procedure (using different frames $f$ and $F$).

### 2.5.1  Procedure: Innermost motion

| Steps to obtain $C^{v-f}(t)$ | In-text ref. |
|---|---|
| 1) Acquire: $\mathbf{v}_Z(\tau)$, $\mathbf{v}_{XZ}(\tau)$, $v_Z^f(\tau)$, $v_{XZ}^f(\tau)$ | 2.4.1, 2.4.2 |
| 2) Calculate: $\Omega_\tau^f$ from $v_Z^f(\tau)$, $v_{XZ}^f(\tau)$ | (34), (35) |
| 3) Calculate: $\mathbf{v}_Z^{-f}(\tau)$, $\mathbf{v}_{XZ}^{-f}(\tau)$ from $\Omega_\tau^f$, $\mathbf{v}_Z(\tau)$, $\mathbf{v}_{XZ}(\tau)$ | (38) |
| 4) Calculate: $\mathbf{v}_X^{-f}(\tau)$, $\mathbf{v}_Y^{-f}(\tau)$ from $\mathbf{v}_Z^{-f}(\tau)$, $\mathbf{v}_{XZ}^{-f}(\tau)$ | (34), (35), (37) |
| 5) Calculate: | (40) |

| | | |
|---|---|---|
| | $\Omega_{\tau,t+\tau}^{\mathbf{v}-f}$ from $\mathbf{v}_X^{-f}(\tau)$, $\mathbf{v}_Y^{-f}(\tau)$, $\mathbf{v}_Z^{-f}(\tau)$ (only $\beta_{\tau,t+\tau}^{\mathbf{v}-f}$ required) | |
| 6) | Calculate: $C^{\mathbf{v}-f}(t) = \left\langle D_{00}^2(\Omega_{\tau,t+\tau}^{\mathbf{v}-f}) \right\rangle_\tau$ from $\Omega_{\tau,t+\tau}^{\mathbf{v}-f}$ | Table 2 |

### 2.5.2 Procedure: Intermediate motion

| | | |
|---|---|---|
| *Steps to obtain $C^{\mathbf{v}:f-F}(t)$* | | *In-text ref.* |
| 1) | Acquire: $\mathbf{v}_Z(\tau)$, $\mathbf{v}_{XZ}(\tau)$, $v_Z^f(\tau)$, $v_{XZ}^f(\tau)$, $v_Z^F(\tau)$, $v_{XZ}^F(\tau)$ | 2.4.1, 2.4.2 |
| 2) | Calculate $A_p^{\mathbf{v}-f}$ | |
| | a. Calculate: $\Omega_\tau^f$ from $v_Z^f(\tau)$, $v_{XZ}^f(\tau)$ | (34), (35) |
| | b. Calculate: $\mathbf{v}_Z^{-f}(\tau)$, $\mathbf{v}_{XZ}^{-f}(\tau)$ from $\Omega_\tau^f$, $\mathbf{v}_Z(\tau)$, $\mathbf{v}_{XZ}(\tau)$ | (38) |
| | c. Calculate: $\mathbf{v}_X^{-f}(\tau)$, $\mathbf{v}_Y^{-f}(\tau)$ from $\mathbf{v}_Z^{-f}(\tau)$, $\mathbf{v}_{XZ}^{-f}(\tau)$ | (34), (35), (37) |
| | d. Calculate: $\Omega_{\tau,t+\tau}^{\mathbf{v}-f}$ from $\mathbf{v}_X^{-f}(\tau)$, $\mathbf{v}_Y^{-f}(\tau)$, $\mathbf{v}_Z^{-f}(\tau)$ ($\beta_{\tau,t+\tau}^{\mathbf{v}-f}, \gamma_{\tau,t+\tau}^{\mathbf{v}-f}$ required) | (40), (41) |
| | e. Calculate $A_p^{\mathbf{v}-f} = \lim_{t\to\infty} \left\langle D_{0p}^2(\Omega_{\tau,t+\tau}^{\mathbf{v}-f}) \right\rangle_\tau$ from $\Omega_{\tau,t+\tau}^{\mathbf{v}-f}$ | Table 2 |
| 3) | Calculate: $\Omega_\tau^f$, $\Omega_\tau^F$ from $v_Z^f(\tau)$, $v_{XZ}^f(\tau)$, $v_Z^F(\tau)$, $v_{XZ}^F(\tau)$ | (34), (35) |
| 4) | Calculate: $\mathbf{v}_Z^{-f}(\tau), \mathbf{v}_{XZ}^{-f}(\tau), \mathbf{v}_Z^{-F}(\tau), \mathbf{v}_{XZ}^{-F}(\tau), v_Z^{f-F}(\tau), v_{XZ}^{f-F}(\tau)$ from $\mathbf{v}_Z(\tau), \mathbf{v}_{XZ}(\tau), v_Z^f(\tau), v_{XZ}^f(\tau), \Omega_\tau^f, \Omega_\tau^F$ | (38) |
| 5) | Calculate: $\mathbf{v}_X^{-F}(\tau)$, $\mathbf{v}_Y^{-F}(\tau)$ from $\mathbf{v}_Z^{-F}(\tau)$, $\mathbf{v}_{XZ}^{-F}(\tau)$ | (34), (35), (37) |
| 6) | Calculate: $\Omega_\tau^{f-F}$ from $v_Z^{f-F}(\tau)$, $v_{XZ}^{f-F}(\tau)$ | (34), (35) |
| 7) | Calculate: $\Omega_{\tau,t+\tau}^{\mathbf{v}:f-F}$ from $\Omega_\tau^{f-F}$, $\mathbf{v}_X^{-F}(\tau)$, $\mathbf{v}_Y^{-F}(\tau)$, $\mathbf{v}_Z^{-F}(\tau)$, $\mathbf{v}_Z^{-f}(\tau)$ ($\alpha_{\tau,t+\tau}^{\mathbf{v}:f-F}, \beta_{\tau,t+\tau}^{\mathbf{v}:f-F}$ required) | (46) |
| 8) | Calculate: $\left\langle D_{p0}^2(\Omega_{\tau,t+\tau}^{\mathbf{v}:f-F}) \right\rangle_\tau$ from $\Omega_{\tau,t+\tau}^{\mathbf{v}:f-F}$ | Table 2 |
| 9) | Calculate: $C^{\mathbf{v}:f-F}(t) = \frac{1}{A_0^{\mathbf{v}-f}} \sum_{p=-2}^{2} \left\langle D_{p0}^2(\Omega_{\tau,t+\tau}^{\mathbf{v}:f-F}) \right\rangle_\tau A_p^{\mathbf{v}-f}$ from $\left\langle D_{p0}^2(\Omega_{\tau,t+\tau}^{\mathbf{v}:f-F}) \right\rangle_\tau$, $A_p^{\mathbf{v}-f}$ | (22) |

### 2.5.3 Procedure: Intermediate motion (Symmetric)

| Steps to obtain $C^{v:f-F}(t)$ | In-text ref. |
|---|---|
| 1) Acquire: $\mathbf{v}_Z(\tau)$, $\mathbf{v}_{XZ}(\tau)$, $v_Z^f(\tau)$, $v_{XZ}^f(\tau)$, $v_Z^F(\tau)$, $v_{XZ}^F(\tau)$ | 2.4.1, 2.4.2 |
| 2) Calculate $A_p^{-f}$ | |
|    a. Calculate: $\Omega_\tau^f$ from $v_Z^f(\tau)$, $v_{XZ}^f(\tau)$ | (34), (35) |
|    b. Calculate: $\mathbf{v}_Z^{-f}(\tau)$, $\mathbf{v}_{XZ}^{-f}(\tau)$ from $\Omega_\tau^f$, $\mathbf{v}_Z(\tau)$, $\mathbf{v}_{XZ}(\tau)$ | (38) |
|    c. Calculate: $\Omega_\tau^{v-f}$ from $\mathbf{v}_Z^{-f}(\tau)$, $\mathbf{v}_{XZ}^{-f}(\tau)$ | (34), (35) |
|    d. Calculate $A_p^{-f} = \left\langle D_{0p}^2(\Omega_\tau^{v-f}) \right\rangle_\tau$ from $\Omega_\tau^{v-f}$ | Table 2 |
| 3) Calculate: $\mathbf{v}_Z^{\text{resid}-f}$ from $A_p^{-f} = \left\langle D_{0p}^2(\Omega_\tau^{v-f}) \right\rangle_\tau$ | Appendix 1 |
| 4) Calculate: $\Omega_\tau^{f-F}$ from $v_Z^{f-F}(\tau)$, $v_{XZ}^{f-F}(\tau)$ | (34), (35) |
| 5) Calculate: $\mathbf{v}_Z^{\text{sym}}(\tau)$ from $\Omega_\tau^{f-F}$, $\mathbf{v}_Z^{\text{resid}-f}$ | (48) |
| 6) Calculate: $\Omega_{\tau,t+\tau}^{\text{sym}-F}$ from $\mathbf{v}_Z^{\text{sym}}(\tau)$ (only $\beta_{\tau,t+\tau}^{\text{sym}-F}$ required) | (49) |
| 7) Calculate: $C^{v:f-F}(t) = \left\langle D_{00}^2(\Omega_{\tau,t+\tau}^{\text{sym}-F}) \right\rangle_\tau$ from $\Omega_{\tau,t+\tau}^{\text{sym}-F}$ | Table 2 |

### 2.5.4 Procedure: Outer motion

The procedure to obtain $C^{v:F}(t)$ is essentially the same as the previous procedure, except no alignment of an outer frame occurs (denoted in the previous section as *F*).

### 2.5.5 Implementing the calculations

Although we have outlined the steps required to obtain correlation functions $C^{v-f}(t)$, $C^{v:f-F}(t)$, and $C^{v:F}(t)$, practical implementation still remains a challenge, especially where we must calculate terms for all *pairs* of time points (that is, all terms $\Omega_{\tau,t+\tau}^{XX}$, and the resulting correlation functions), since it becomes too time consuming to perform loops over all required dot products and matrix multiplications. However, the various linear algebra operations can be re-cast as summations of products of components of the various vectors. In Appendix 2, we show that all the angular correlation functions required can be calculated as a sum of linear correlation functions, and furthermore in Appendix 3, show that these linear correlation functions may be obtained via fast-Fourier transform [48], yielding significant reduction in computational time. Components of residual tensors, $A_p^X$ may also be calculated as a sum of products of variances and covariances.

## 3   Results and Discussion

In the previous sections, we have discussed under what conditions a multi-component motion may be separated into parts, yielding individual correlation functions that may be multiplied together to yield the total correlation function (sections 2.2, 2.3), and how one goes about calculating those correlation functions explicitly from a simulated trajectory (sections 2.4, 2.5). Here we illustrate with examples how the resulting correlation functions actually behave, and show limitations in the proposed approach.

### 3.1   Example Calculations

#### 3.1.1   *Four motions with ideal behavior*

As a first example, we generate an "ideal" trajectory, where four motions are separated using frame analysis. The trajectory is synthetic so that we know that motions are statistically independent, with mono-exponential decay (we generate Poisson distributions for the time dependence of hopping), and timescale separated. Furthermore, we assume our frame definitions can perfectly extract the real motion of each frame (in practice, a frame is defined by coordinates in the MD trajectory and may inadvertently incorporate undesired motions). Then, motions included in the trajectory are intended to mimic motion of an H–C bond in a methyl group, which is in turn moving due to reorientation of its C–C bond and due to the molecule tumbling in solution. The motions are as follows:

1) Fast wobbling in a cone: This motion will cause a bond vector to move randomly from various orientations within a cone, with an opening angle of 15° (z-axis to furthest angle away from *z*). There is no correlation between time points, so that the correlation time is effectively 0 (or, significantly shorter than the time step, taken to be 5 ps). This motion is similar to librational motion in simulations, where the correlation time is often shorter than the recorded time-step.

2) Symmetric three-site hopping: We take a three-step rotation around an axis (120° steps). This is similar to what is expected for a methyl rotation. However, we note for a methyl rotation that the angle between the bond and rotation axis is ~109.47°, yielding $S^2$ of 1/9. Such a motion dominates the overall correlation function, making the influence of slower motions strongly reduced, and hard to see. To avoid masking other motions, we set the angle between bond and rotation axis to a less-physical value of 150° ($S^2$=0.39). We set the mean time between jumps to 100 ps (leading to $\tau_{met}$ = 66.7 ps). Generation of the trajectory is performed using a random binomial, where jumping between orientations occurs with probability $\Delta t / \langle \tau \rangle$, with $\Delta t$ being the time step and $\langle \tau \rangle$ being the mean time between jumps. The direction of the hop is chosen with 50/50 probability.

3) Two-site hop: We use a two-site hop between two orientations, separated by an angle of 45°. The mean time between hops is 1 ns, resulting in $\tau_{hop}$ =500 ps. Trajectory generation is performed the same way as methyl rotation (without selecting a direction, since we always go to the other site).

4) Isotropic tumbling: Finally, we apply overall isotropic tumbling. This is achieved by rotating a vector 0.5° away from its current direction at every time step, but with a random angle (achieving a random walk around a sphere). Using a 5 ps timestep, this results in a correlation time of ~42 ns.

The four motions are separated by three intermediate frames, referred to as: methyl-aligned frame (ME), C–C aligned frame (CC), molecule-aligned frame (MO), and the lab frame (LF). Then, librational motion is the result of fast wobbling motion of the H–C bond within the ME frame ($\Omega_\tau^{PAS,ME}$), methyl rotation is the result of motion of the ME frame within the C–C frame ($\Omega_\tau^{ME,CC}$), hop motion is the result of motion of the CC frame within the MO frame ($\Omega_\tau^{CC,MO}$), and finally tumbling is the result of the MO frame within the lab frame ($\Omega_\tau^{MO,LF}$). Then, the total motion is the result of all individual motions:

$$\mathbf{v}_Z^{LF}(\tau) = \mathbf{R}_{ZYZ}(\Omega_\tau^{MO,LF}) \cdot \mathbf{R}_{ZYZ}(\Omega_\tau^{CC,MO}) \cdot \mathbf{R}_{ZYZ}(\Omega_\tau^{ME,CC}) \cdot \mathbf{R}_{ZYZ}(\Omega_\tau^{PAS,ME}) \cdot [0,0,1]'. \qquad (50)$$

For each calculation, we use a step size of 5 ps, and include $10^6$ time points, for a total length of 5 µs. When plotting correlation functions, we omit the second half, since these points are calculated with fewer pairs of time points and thus become noisy.

Results are shown in Fig. 5. In Fig. 5A, we show the correlation function directly calculated for the total motion (red, solid) compared to the correlation function obtained as a product of correlation functions of the individual motions (black, dotted). We see that the two correlation functions are in excellent agreement, confirming successful frame decomposition. In Fig. 5B, we see the correlation functions of the individual motions. These have been constructed according to eq. (22). In Fig. 5C, we show components of the correlation function for librational motion. Since librations are extracted from motion within the first frame, they are obtained based only on $\langle D^2_{00}(\Omega^{v-f}_{\tau,t+\tau})\rangle_\tau$ (Fig. 5B, black), whereas other terms are only shown only for comparison to the same terms for other motions. In subsequent sections, all terms $\langle D^2_{p0}(\Omega^{v:f-F}_{\tau,t+\tau})\rangle_\tau$ are required for calculating the correlation function for that motion (Fig. 5D-F, left, color). Then, the correlation function for each motion is the sum of the product $A^{v-f}_p \langle D^2_{p0}(\Omega^{v:f-F}_{\tau,t+\tau})\rangle_\tau$ (Fig. 5D-F, left, black plots this sum). The $A^{v-f}_p$ are given by $\lim_{t\to\infty}\langle D^2_{0p}(\Omega^{v-f}_{\tau,t+\tau})\rangle_\tau$, where the evolution of the $\langle D^2_{0p}(\Omega^{v-f}_{\tau,t+\tau})\rangle_\tau$ is shown to the right of each section (Fig. 5D-F). Then, the assumption of timescale separation is satisfied if the $\langle D^2_{0p}(\Omega^{v-f}_{\tau,t+\tau})\rangle_\tau$ have reached their equilibrium values before significant evolution of the terms $\langle D^2_{p0}(\Omega^{v:f-F}_{\tau,t+\tau})\rangle_\tau$ have occurred (we do not need the full correlation functions shown in Fig. 5D-F, right, to calculate $C^{v:f-F}(t)$, only the equilibrium values, $A^{v-f}_p$, but show the evolution to demonstrate that they have equilibrated). In our example, the motions are deliberately timescale separated, but if they are not, then change in residual tensor shape due to motion in frame *f* may prevent accurate calculation of the correlation function of motion of frame *f*. That is, because the residual tensor is not equilibrated, there is no "correct" tensor shape to which we can apply the motion of frame *f*.

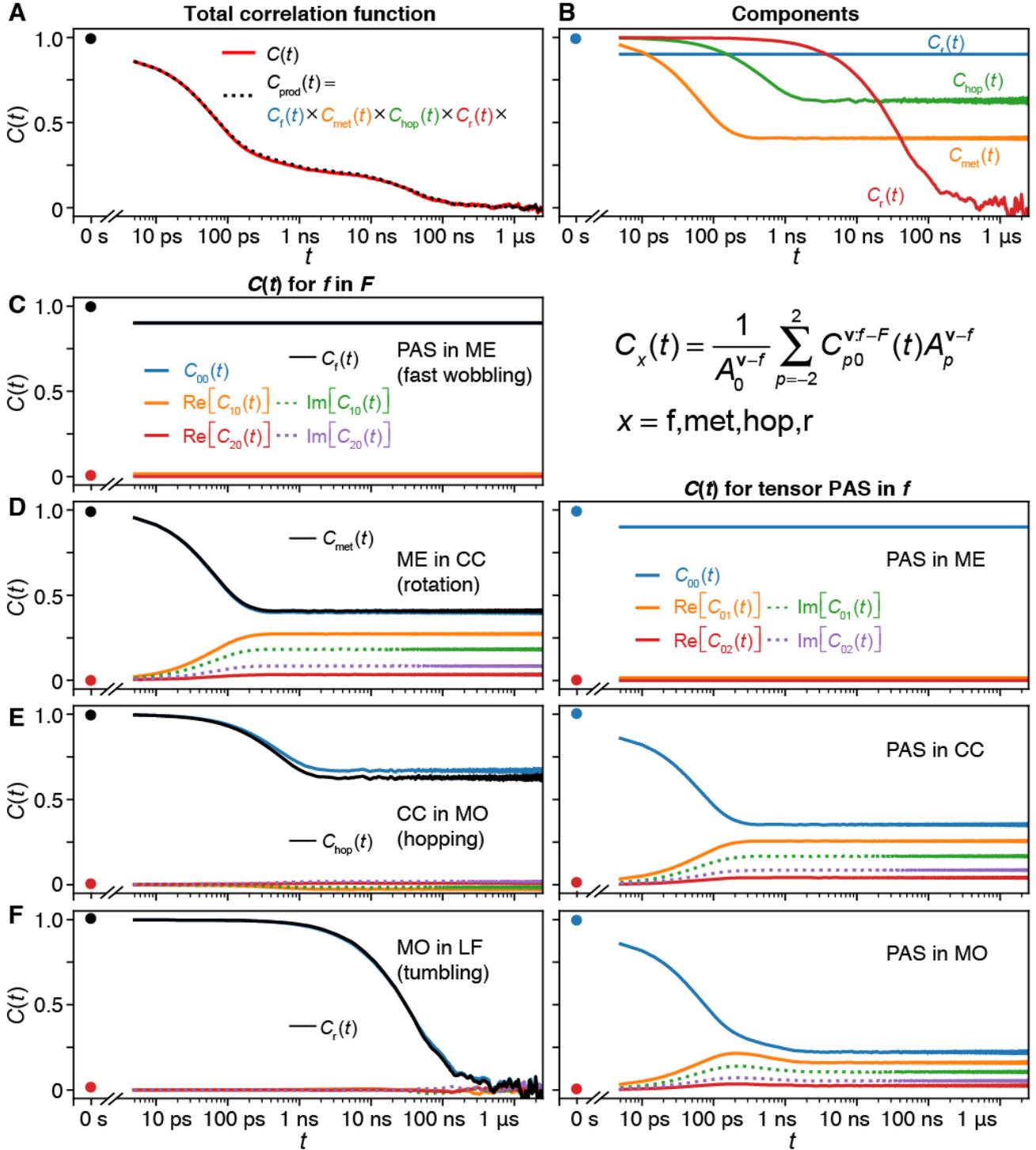

**Fig. 5.** Four motions separated into components. **A** shows the directly calculated total correlation function (red), and the product of all individual correlation functions (black, dotted). **B** shows the individual correlation functions for fast motion, symmetric three-site hopping, two-site hopping, and tumbling. **C–F** show the components used to construct each of the individual correlation functions. At the left we show the components of the correlation function for motion of frame $f$ in frame $F$. At the right, we show the correlation functions for motion of the PAS in $f$. The total correlation function is obtained by multiplying each of the components of the correlation functions from motion of $f$ in $F$ with the corresponding component of the PAS in $f$ correlation function at infinite time (eq. (22)). Finally, one takes the sum of the products for each of the five components, shown in black on the left panel of **C–F**. Note we use $C_{0p}(t)$ in the figure as shorthand for $\langle D^2_{0p}(\Omega^{v-F}_{\tau,t+\tau})\rangle_\tau$ or $\langle D^2_{0p}(\Omega^{v:f-F}_{\tau,t+\tau})\rangle_\tau$.

*3.1.2 Two motions with non-ideal behavior*

In the subsequent calculations, we only define two motions (usually, three-site hopping and two-site hopping, as in the previous example), separated by a single frame (C–C aligned frame). We use these to show a variety of mechanisms that may cause failure of the frame analysis (we consider failure to be when the product of the individual correlation functions does not yield the total correlation function).

The first mechanism of failure is simply a poor choice of frame. What occurs if we want to separate three-site hopping of the H–C bond from hopping motion of the C–C bond, but the chosen frame is not actually correlated with the overall H–C dynamics? In Fig. 6A (top), we define the frame to be the C–C bond direction, and find that the product of correlation functions accurately reproduces directly calculated correlation function. If the C–C bond motion does not influence the H–C bond motion, the correlation functions do not agree, as shown in Fig. 6A (bottom). In this case, the frame definition is relatively obvious–clearly if we correctly select the C–C bond, the methyl H–C group should move with it (in practice, selecting the wrong bond might produce this failure). More subtle mechanisms of failure can also occur: suppose we defined a frame via alignment of part of the protein backbone to a reference structure: this could be an effective frame choice for motion of backbone H–N bonds within a structured region of the protein, separating local motion from collective motion of the given secondary structure element (e.g. α-helix, β-sheet). However, if we analyzed bonds in a disordered region of the protein with alignment to a reference structure, the lack of a well-defined structure for that region would likely produce a frame whose motion is not well correlated with the motion of the H–N bonds.

The next source of failure is lack of timescale separation. In Fig. 6B (top), we again consider the three-site and two-site hops, where the three-site rotation has a correlation time of 33 ps and the two-site hop has a correlation time of 5 ns. Then, the residual tensor of the three-site hop is fully equilibrated before two-site hopping occurs. On the other hand, we can change the correlation time of three-site hopping to 33 ns and the two-site hopping to 50 ps, so that the residual tensor of the three-site hop is not equilibrated when two-site hopping occurs, leading to the disagreement observed in Fig. 6B (bottom). Note that in this case, the problem could be resolved by changing the frame such that the faster motion (two-site hopping) is the motion within frame *f*, and the slower motion (three-site rotation) is the motion of frame *f*. However, defining a frame that swaps the order of these motions is not trivial.

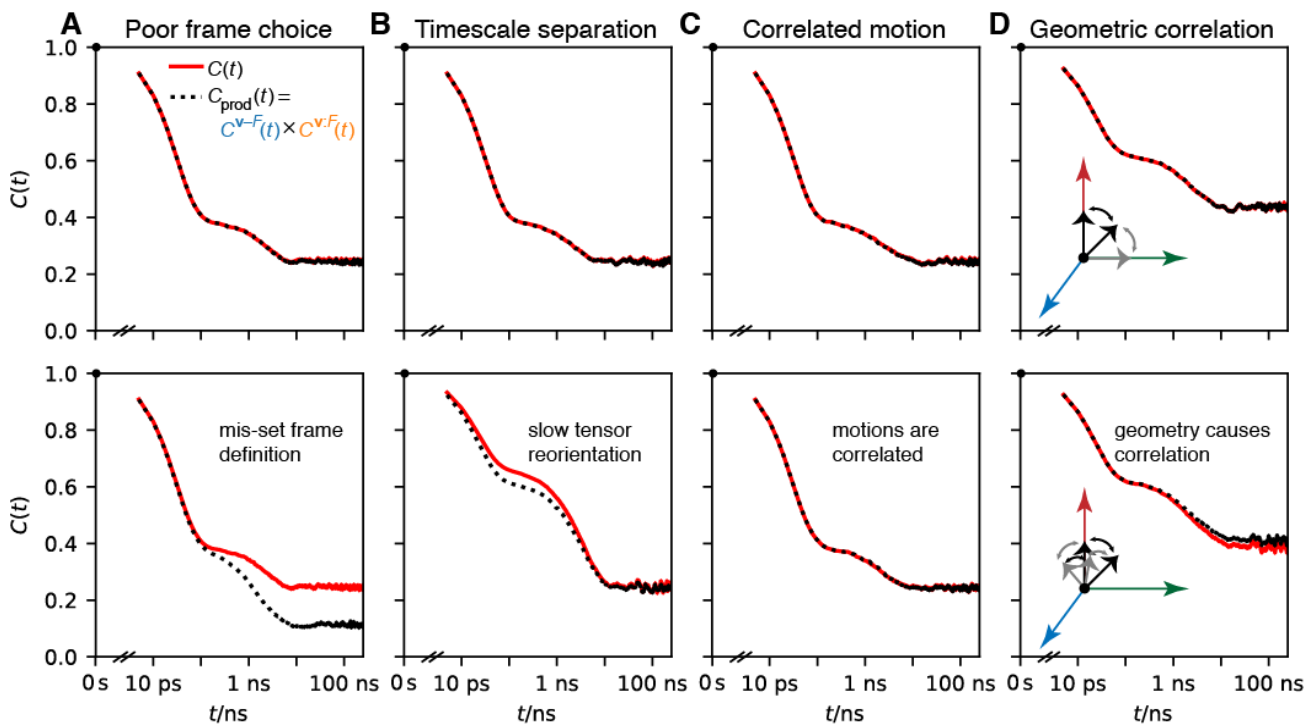

**Fig. 6.** Modes of frame analysis failure. In each subplot, we generate a trajectory based on two simple motions. In **A**-**C** we use a three-site symmetric rotation (methyl hopping), with an opening angle of 150° and correlation time of 33 ps as the motion in frame *f* and a two-site hop with an angle of 45° and correlation time of 5 ns for the motion of frame *f* (unless otherwise noted). In **A**, we show the separation with frame *f* correctly defined (top), and incorrectly defined (bottom) such that motion of frame *f* is a two-site hop, but it is uncorrelated with the total motion. In **B**, we show separation with frame *f*, where the inner motion is faster than the outer motion (top, 33 ps vs. 50 ns), and where the inner motion is much slower than the outer motion (33 ns vs. 50 ps), such that tensor reorientation due to motion in frame *f* is slower than motion of frame *f*. In **C**, we show the influence of correlation of motion between the frames, where at top, the motion is uncorrelated. In the bottom plot, we only allow the two-site hop to occur when the symmetric rotation is in the first position (thus making the motions are correlated). In **D**, we replace the three-site rotation with a two-site hop with an angle of 45° and correlation time of 50 ps. At top, both two-site hops result in rotation around the *y*-axis, but at bottom, the faster hop rotates about the *y*-axis but the slower hop rotates about the *x*-axis, yielding a geometric correlation.

In Fig. 6C (top), we again start with three-site rotation and two-site hopping, and test how correlation of the motions will affect our results. In the top plot, motions are uncorrelated but in the bottom plot, we allow the two-site hop to occur only when the 3-site hopping is one of the three positions, thus introducing strong correlation of the motions. However, no disagreement in the correlation functions results, indicating that statistical independence of the terms is retained. In contrast, Fig. 6D compares two trajectories, both with two-site hopping, but in Fig. 6D(top), the two hops both occur about the *x*-axis (so that the two 45° hops may lead to a net reorientation of 90°), but in Fig. 6D(bottom) one hop occurs about the *x*-axis and the other about the *y*-axis. Although no correlation is introduced between the hops, and the motions are well timescale-separated, the correlation functions do not agree exactly. Here, the position of the tensor due to hoping about the *x*-axis changes the influence that a hop about the *y*-axis has on the decay of the correlation function. This geometric correlation is not fully captured by the frame analysis, and so leads to minor deviations of the directly calculated correlation function and product of correlation functions. Then, one should keep in mind that "statistical dependence" is very specifically defined as when the terms in eq. (16) become zero, and is not always related in an obvious way to correlation of the motions (although sometimes the behavior will be as expected).

*3.1.3 Methyl dynamics from an MD trajectory*

While some of the behavior resulting from correlation of motions is somewhat unexpected, overall we find excellent performance of the frame analysis, with only severe mis-definition of the frames or very poor timescale separation leading to large deviations. However, the behavior here is ideal, where we are guaranteed that motions are uncorrelated and frames are well defined. Therefore, we test the frame analysis performance on a real MD trajectory. We use a 1 μs simulation of HET-s(218-289) fibrils (AMBER ff99SB*-ILDN [49] with methyl rotation barrier correction [5,6], run with GROMACS [50], trajectory courtesy of Kai Zumpfe), and examine the methyl dynamics of Ile254 (we examine the middle molecule of a fibril having 5 HET-s copies). In the first example, we separate the total motion using the Cγ–Cδ bond as the first frame, the Cβ–Cγ bond as the second frame, and the Cα–Cβ bond as the third frame (these define $v_z^f(\tau)$ for each frame, although note that for each frame, we also define $v_{xz}^f(\tau)$, using the next bond towards the backbone to avoid ambiguity in the frame orientation). Fig. 7A shows the correlation functions for rotation around each bond (this set of frame definitions also cause some tensor reorientation due to bond angle fluctuations to be captured by each frame). Note that the fourth plot shows the influence of reorientational motion of the Cα–Cβ bond, capturing primarily backbone dynamics and bond angle fluctuations. In the bottom plot, we see that the product of the individual correlation functions does an excellent job reproducing the total correlation function.

Some very minor deviation, however, can be observed after the very first time point (5 ps). One might consider such deviation negligible, but we note that it appears almost universally when analyzing real MD simulations. To understand this behavior, we point out that correlation functions for all components of the motion have a rather sharp drop at the first time point. This is likely to be fast librational motion, and the disagreement in Fig. 7A(bottom) then results because these librational motions are correlated for the different frames. Therefore, the disagreement may be removed by averaging the frame direction over a number of frames in order to remove the correlated libration of the frames (we use a Gaussian weighting for averaging, with σ=5 ps for the first frame, and 50 ps for the second and third frame. σ is chosen to be longer than the mean correlation time for each correlation function). This is shown in Fig. 7B, where the sharp drop in the individual correlation functions is partially or fully removed and the product of correlation functions is in better agreement with the directly calculated correlation function (bottom).

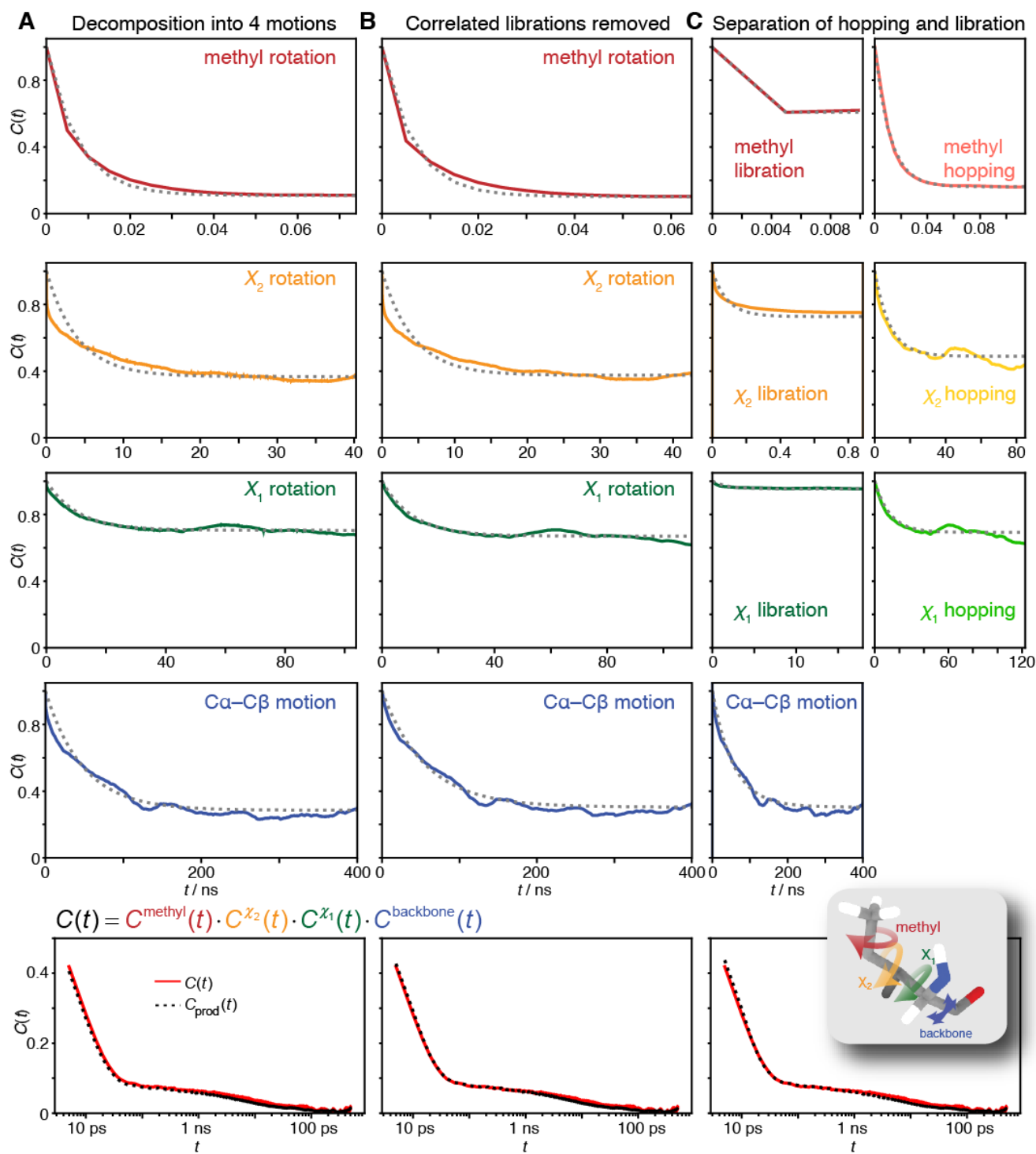

**Fig. 7.** Decomposition of methyl group motion in an isoleucine (HET-s 254Ile). In **A**, frames are defined based on alignment of the Cγ–Cδ bond, Cβ–Cγ bond, and Cα–Cβ bonds, so that motion is approximately separated into methyl rotation (top), $\chi_2$ rotation (middle top), $\chi_1$ rotation (middle), and Cα–Cβ motion (middle bottom) noting that bond libration and deviation from ideal geometry also contribute to decay of the correlation functions. The directly calculated correlation function (red, solid) is compared to the product of correlation function (black, dotted) in the bottom plot. In **B**, fast libration motion is averaged out in the frame definitions, yielding a slightly improved agreement of the directly calculated correlation function and the product of individual correlation functions. In **C**, rotational motions (methyl, $\chi_2$, $\chi_1$) are split into librational contributions and 3-site hopping contributions (orientations fixed to three positions corresponding to the energy minima). In each plot, a grey dotted line plots a decaying exponential having the same mean correlation time and order parameter as the real correlation function, highlighting mono- and multiexponential behavior.

Although we have successfully separated the motion of H–C bonds in methyl groups into four components, we notice that some of the components of the motion are clearly multi-exponential (grey, dashed lines in Fig. 7 indicate a mono-exponentially decaying function with the same order parameter and mean correlation time as the correlation function calculated from MD). This is particularly clear for methyl and $\chi_2$ rotation in Fig. 7B. The multi-exponentiality indicates that at least two distinct modes of motion manifest in these correlation functions. In fact, there are several possible motions, including hopping between the three energy minima for the $\chi_2$ rotation, smaller amplitude rotations about the Cβ–Cγ bond, and distortions of the local geometry. Then, in a last step, we introduce new frames that only capture hops between three sites of the methyl, $\chi_1$, and $\chi_2$ rotation, with results in Fig. 7C. Methyl, $\chi_1$, and $\chi_2$ rotations are split into two correlation functions, one capturing 3-site hopping (Fig. 7C, right) and one capturing the remaining motion (primarily rotations and bond angle fluctuations, Fig. 7C, left). The former correlation function is nearly monoexponential, and significantly slower than the latter function, due to the higher activation energy required for jumping between the three energy minima. Then, such a decomposition could be used in concert with detector analysis of methyl dynamics (e.g. [37]) to determine in which detector window each of these seven motions occurs, thus allowing one to better interpret the detector responses.

*3.1.4 Protein backbone dynamics with detector analysis*

Finally, we investigate how to separate motions influencing backbone dynamics in a helical protein, the $Y_1$ G-coupled protein receptor (GPCR) [51], using an MD simulation provided by Vogel et al. [44]. In this case, we use three frames to separate backbone H–N dynamics into four components. The first frame is defined by alignment of each peptide plane to a reference structure (peptide plane defined by the H, N, and Cα of the given residue and C', O, and Cα of the previous residue). Then, this frame separates H–N librational motion within the peptide plane from all other motions. The next frame is defined by alignment of α-helices (Cα alignment) to a reference structure (we split α-helices into intracellular and extracellular regions, according to Table 3 in Vogel et al. [44]), separating peptide plane motion from concerted helix motion. Finally, we define one frame that aligns the whole protein (all Cα in helices are used) to separate α-helix motions from overall motion of the protein within the membrane. We analyze the resulting correlation functions using detector analysis [35], and show results for the whole protein. In Fig. 8A-B, we use detectors to compare amplitude of motion for a range of timescales, obtained for the directly calculated total correlation functions and the correlation functions obtained as a product of the individual motions. Good agreement in all detector windows indicates that the ROMANCE approach has been successful in separating the various motions (caution should be taken in interpretation for residues and timescales with poorer agreement). Note that the detector sensitivity for $\rho_6$ extends to infinitely long correlation times (Fig. 8A), indicating that it captures non-decaying components of the correlation function (i.e. $S^2$), but also slowly decaying components, which have not decayed within the length of the simulated trajectory (non-decaying and slowly decaying components of the correlation function are indistinguishable in a finite trajectory). In Fig. 8C, we map the total amplitude of reorientational motion (defined by $1-S^2$) onto a frame of the trajectory and in Fig. 8D-G we separate the total motion by timescale and by type of motion.

In Fig. 8D, we plot H–N motion within the peptide plane frame (librational motion), where we see that this motion is very uniform and only appearing at the shortest correlation times (<900 ps, note that this trajectory was stored every 100 ps, so that one cannot easily distinguish correlation time below this timescale). Peptide plane motion (Fig. 8E), on the other hand, depends on the location within the α-helix, with greater amplitude occurring near the ends of helices and in between intra- and extracellular regions. Fig. 8F shows overall motion of the helices (divided into intra- and extracellular regions), where one finds that extracellular dynamics are significantly increased in the microsecond regime. Finally Fig. 8G shows the overall motion of the protein in the membrane, which occurs in the low microsecond regime. Note that Fig. 8G illustrates a critical feature of the ROMANCE analysis: the overall motion of $Y_1$ GPCR is, by definition, identical for all residues. However, this does not result in the same detector response for all residues. For example, helix 4 (far right side of molecule) has noticeably smaller detector responses in the µs-regime ($\rho_4$-$\rho_5$) for the overall motion than other helices. This is because the overall motion is predominantly a rotation around the membrane normal, whereas residual H–N dipole couplings in this helix are approximately parallel to the membrane normal. Thus overall motion induces very little decay of the correlation function on helix 4, and also very little relaxation (the influence of the relative directions of the motion and the residual tensor is a consequence of eq. (22)). This behavior can be contrasted with the analysis of this trajectory by Vogel et al. [44], where the overall motion of each helix was characterized without considering the relative tensor directions and directions of the motion. The advantage of the approach from Vogel et al. is that information about motion of the helix itself is not convoluted by the influence of the direction and shape of the residual H–N dipole couplings, however the advantage of the ROMANCE approach is that we may determine how individual motions influence experimental relaxation behavior, considering complex factors such as the relative direction of the residual tensors and the motion itself.

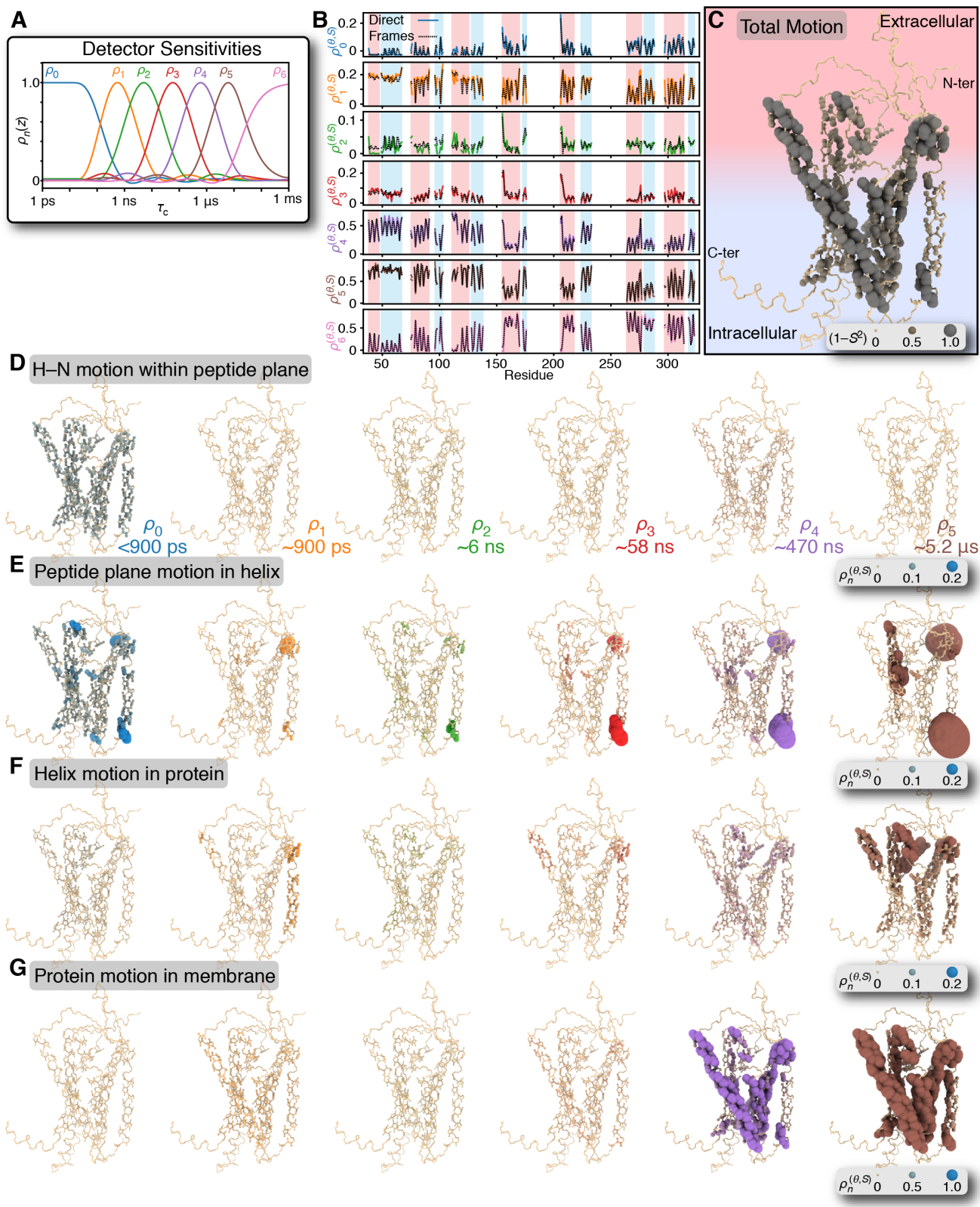

**Fig. 8.** Combined Detector and ROMANCE analysis of H–N motion in $Y_1$ GPCR [44]. H–N motion is broken into four components: H–N librational motion within the peptide plane motion, peptide plane motion within the helix section (helices separated into intracellular and extracellular regions), helix motion within the protein, and protein motion within the membrane (i.e. lab frame). **A** shows the detector sensitivities and **B** shows detector analysis of the directly calculated correlation function (color) and the correlation function obtained as a product of the correlation functions derived for the individual motions (black, dashed). **C** shows the total amplitude of H–N motion ($1-S^2$) mapped onto $Y_1$ (radius and color intensity for all atoms in the corresponding peptide plane are proportional to the amplitude). **D**-**G** shows the amplitude of motion separated by type of motion (via ROMANCE) and by timescale (via detector analysis). Approximate timescales are indicated on the 3D plots in **D**. Amplitude scale indicated in the lower right of each plot. Molecule plots created with ChimeraX [41].

# 4    Conclusions

In this study, we have re-visited the theoretical basis for separation of the correlation function of reorientational motion into a product of correlation functions resulting from separate contributions to the reorientational motion. Subsequently, we have demonstrated how to calculate the individual correlation functions based on coordinates in an MD simulation, explicitly taking account of the shape of residual tensors resulting from inner motions and the relative direction of those tensors and the direction of the outer motion. The result is a highly accurate decomposition of the total correlation function as a product of correlation functions of individual motions. The ROMANCE analysis may then be used in combination with detector analysis, to first perform quantitative comparison of timescale-specific dynamics in experiment and simulation [10,38,39], and to then interpret the experimental results using timescale-sensitive analysis of individual motions in the MD trajectory. Such a tool will be valuable in better analyzing and interpreting experimental NMR relaxation data, and the theory for achieving this separation may serve as the basis for future techniques joining NMR relaxation data and MD simulation.

# 5    Acknowledgements

This project was funded by the Deutsche Forschungsgemeinschaft (DFG), grant SM 576/1-1. Thanks to Kai Zumpfe and Alexander Vogel for providing MD simulations used in Fig. 7 and Fig. 8, respectively.

# 6    Code Availability

Python scripts, including a beta-version of pyDIFRATE for detector analysis and a trajectory of HET-s residue 254, have been made available on GitHub, and can be used to reproduce Fig. 5–Fig. 7 [52].

# Appendix

## 1 Obtaining δ, η, Ω for an interaction tensor

If we have a residual tensor, with components $A_p$, we sometimes need to determine its anisotropy, asymmetry, and Euler angles, for example, as required for eq. (48). One approach is to first transform the spherical tensor into Cartesian coordinates:

$$\begin{bmatrix} A_{XX} \\ A_{XY} \\ A_{XZ} \\ A_{YY} \\ A_{YZ} \end{bmatrix} = \begin{bmatrix} 1/2 & 0 & -\sqrt{1/6} & 0 & 1/2 \\ i/2 & 0 & 0 & 0 & -i/2 \\ 0 & 1/2 & 0 & -1/2 & 0 \\ -1/2 & 0 & -\sqrt{1/6} & 0 & -1/2 \\ 0 & i/2 & 0 & i/2 & 0 \end{bmatrix} \begin{bmatrix} A_{-2} \\ A_{-1} \\ A_0 \\ A_1 \\ A_2 \end{bmatrix} \tag{A1}$$

Noting that the Cartesian matrix is symmetric, it is possible to find a matrix, **U**, which diagonalizes the matrix (Eigenvalue decomposition).

$$\begin{bmatrix} A_{XX}^{PAS} & 0 & 0 \\ 0 & A_{YY}^{PAS} & 0 \\ 0 & 0 & A_{ZZ}^{PAS} \end{bmatrix} = \mathbf{U}' \cdot \begin{bmatrix} A_{XX} & A_{XY} & A_{XZ} \\ A_{XY} & A_{YY} & A_{YZ} \\ A_{XZ} & A_{YZ} & -(A_{XX}+A_{YY}) \end{bmatrix} \cdot \mathbf{U} \tag{A2}$$

Caution should be taken to ensure that the resulting diagonal matrix satisfies $\left|A_{ZZ}^{PAS}\right| \geq \left|A_{XX}^{PAS}\right| \geq \left|A_{YY}^{PAS}\right|$ (this may be achieved by re-ordering the columns of **U**). Then, we may obtain δ and η the usual way, considering that the tensor is traceless:

$$\begin{aligned} \delta &= A_{ZZ}^{PAS} \\ \eta &= \frac{A_{YY}^{PAS} - A_{XX}^{PAS}}{A_{ZZ}^{PAS}} \end{aligned} \tag{A3}$$

**U** is a rotation matrix, and if it is a proper rotation matrix, we may extract the Euler angles from it. To ensure a proper rotation matrix, **U** should have a determinant of 1 (achieved by multiplying **U** by its own determinant, where the determinant is either 1 if **U** is proper or -1 if **U** is improper). Then, the proper rotation matrix may be expressed as a function of the Euler angles (*c* and *s* are shorthand for cosine and sine).

$$\mathbf{U} = \begin{bmatrix} c\gamma c\beta c\alpha - s\gamma s\alpha & c\gamma c\beta s\alpha + s\gamma c\alpha & -c\gamma s\beta \\ -s\gamma c\beta c\alpha - c\gamma s\alpha & -s\gamma c\beta s\alpha + c\gamma c\alpha & s\gamma s\beta \\ s\beta c\alpha & s\beta s\alpha & c\beta \end{bmatrix} \tag{A4}$$

We define *β* to be in the range 0, π, such that its value may be unambiguously extracted from **U**[3,3]. Then, $\sin\beta$ is always non-negative. If $\sin\beta \neq 0$, then $\cos\alpha$ and $\sin\alpha$ ar easily extracted from **U**[3,1] and **U**[3,2], and similarly $\cos\gamma$ and $\sin\gamma$ are extracted from **U**[1,3] and **U**[2,3]. If *β*=0, on the other hand, then *α* and *γ* have the same effect and so cannot be separated. Then, we set *α*=0, and extract $\cos\gamma$ from **U**[1,1] and $\sin\gamma$ from **U**[2,1].

Note that while extraction of Euler angles from **U** is unambiguous, multiple solutions for **U** exist. However, for purposes here, any valid set of Euler angles is sufficient.

## 2  Correlation functions as sums

In the main text, we have shown how to obtain the required correlation functions as either dot products of vectors or products of vectors and rotation matrices. While one may solve the correlation functions from these various formulas, doing so directly may take an excessive amount of computation time. Here, we first show how to re-cast all correlation functions as sums of linear correlation functions (Appendix 2) and additionally show how to rapidly calculate the equilibrium values of those correlation functions (Appendix 3.1) and the correlation functions themselves (Appendix 3.2).

Eqs. (40), (41), (45) and (46) provide the Euler angles required to calculate the various correlation functions, although the notation does not lend itself to simple translation into computer code for practical calculation. Here, we rewrite these equations as sums over terms, rather than dot products of vectors and matrices, which may then more easily be implemented in a computer program.

### 2.1  Correlation functions for motion within frame *f*

#### 2.1.1  Required Euler angles

We begin with the correlation function for motion within frame *f*, which requires just the term $\left\langle D^2_{00}(\Omega^{v-f}_{\tau,t+\tau})\right\rangle_\tau = \left\langle d^2_{00}(\beta^{v-f}_{\tau,t+\tau})\right\rangle_\tau$. Although here we only require the angle $\beta^{v-f}_{\tau,t+\tau}$, we also will need the angles $\alpha^{v-f}_{\tau,t+\tau}$ and $\gamma^{v-f}_{\tau,t+\tau}$ for calculation of the residual tensor, terms $A^{v-f}_p$. Then, from eqs. (40) and (41), we expand the dot product for each element of the vector to obtain the required angles.

$$\cos\beta_{\tau,t+\tau}^{\mathsf{v}-f} = \mathbf{v}_Z^{-f}(\tau) \cdot \mathbf{v}_Z^{-f}(t+\tau)$$

$$= \begin{bmatrix} x_Z^{-f}(\tau) & y_Z^{-f}(\tau) & z_Z^{-f}(\tau) \end{bmatrix} \cdot \begin{bmatrix} x_Z^{-f}(t+\tau) \\ y_Z^{-f}(t+\tau) \\ z_Z^{-f}(t+\tau) \end{bmatrix}$$

$$= \sum_{\zeta=x,y,z} \zeta_Z^{-f}(\tau)\zeta_Z^{-f}(t+\tau)$$

$$\cos\alpha_{\tau,t+\tau}^{\mathsf{v}-f} \sin\beta_{\tau,t+\tau}^{\mathsf{v}-f} = -\sum_{\zeta=x,y,z} \zeta_X^{-f}(t+\tau)\zeta_Z^{-f}(\tau) \cdot \quad (A5)$$

$$\sin\alpha_{\tau,t+\tau}^{\mathsf{v}-f} \sin\beta_{\tau,t+\tau}^{\mathsf{v}-f} = \sum_{\zeta=x,y,z} \zeta_Y^{-f}(t+\tau)\zeta_Z^{-f}(\tau)$$

$$\sin\beta_{\tau,t+\tau}^{\mathsf{v}-f} \cos\gamma_{\tau,t+\tau}^{\mathsf{v}-f} = \sum_{\zeta=x,y,z} \zeta_X^{-f}(\tau)\zeta_Z^{-f}(t+\tau)$$

$$\sin\beta_{\tau,t+\tau}^{\mathsf{v}-f} \sin\gamma_{\tau,t+\tau}^{\mathsf{v}-f} = \sum_{\zeta=x,y,z} \zeta_Y^{-f}(\tau)\zeta_Z^{-f}(t+\tau)$$

where $\zeta_\lambda^{-f}(\tau)$ are elements of the vectors $\mathbf{v}_\lambda^{-f}(\tau) = [x_\lambda^{-f}(\tau), y_\lambda^{-f}(\tau), z_\lambda^{-f}(\tau)]$. We have expanded the first equation, whereas the subsequent equations may be treated analogously

Using the Euler angles above, we may now derive the individual terms required to calculate the various correlation functions. We derive all the terms required when only separation of timescale is present (section 2.2.2). However, in case of isotropic tumbling, then we only require the terms $\langle D_{00}^2(\Omega_{\tau,t+\tau}^{\mathsf{v}-f}) \rangle_\tau$ for the innermost motion, and $\langle D_{00}^2(\Omega_{\tau,t+\tau}^{\mathsf{v}:f-F}) \rangle_\tau$ for the outer motion. Note that we only need the equilibrium values for the terms $\langle D_{0p}^2(\Omega_{\tau,t+\tau}^{\mathsf{v}-f}) \rangle_\tau$ for $p \neq 0$, although we will also show how to also obtain the time-dependence since it useful for observing the evolution of the residual tensor. In case the inner motion has an axis of symmetry that is not parallel to the tensor (section 2.2.3), then we calculate $\langle D_{00}^2(\Omega_{\tau,t+\tau}^{\mathrm{sym}-F}) \rangle_\tau$ instead of $\langle D_{00}^2(\Omega_{\tau,t+\tau}^{\mathsf{v}:f-F}) \rangle_\tau$. We start by calculating terms for motion within frame *f*.

### 2.1.2 Calculating the correlation functions

The correlation function for motion of the interaction within frame *f*, $C^{\mathsf{v}-f}(t)$, depends only on the term $\langle D_{00}^2(\Omega_{\tau,t+\tau}^{\mathsf{v}-f}) \rangle_\tau$, although we will derive all terms $\langle D_{0p}^2(\Omega_{\tau,t+\tau}^{\mathsf{v}-f}) \rangle_\tau$, since their equilibrium values are required for calculating $C^{\mathsf{v}:f-F}(t)$. Starting with the (0,0) component:

$$C_{00}^{v-f}(\tau) = \left\langle D_{00}^2(\Omega_{\tau,t+\tau}^{v-f}) \right\rangle_\tau = \left\langle \frac{3\cos^2\beta_{\tau,t+\tau}^{v-f} - 1}{2} \right\rangle_\tau$$

$$= -\frac{1}{2} + \frac{3}{2}\left\langle \left( \sum_{\zeta=x,y,z} \zeta_Z^{-f}(\tau)\zeta_Z^{-f}(t+\tau) \right)^2 \right\rangle_\tau \qquad (A6)$$

$$= -\frac{1}{2} + \frac{3}{2} \sum_{\zeta=x,y,z}\sum_{\xi=x,y,z} \left\langle \zeta_Z^{-f}(\tau)\xi_Z^{-f}(\tau)\zeta_Z^{-f}(t+\tau)\xi_Z^{-f}(t+\tau) \right\rangle_\tau$$

We have simply inserted the function for $\cos\beta_{\tau,t+\tau}^{v-f}$ from eq. (A5) into the expression for $\left\langle D_{00}^2(\Omega_{\tau,t+\tau}^{v-f}) \right\rangle_\tau$ ( $C_{0p}^{v-f}(t) \equiv \left\langle D_{0p}^2(\Omega_{\tau,t+\tau}^{v-f}) \right\rangle_\tau$ ). The squared term is expanded, noting that an expectation value of a sum is equal to the sum of the expectation values of the individual terms within the sum. The result is a sum of linear correlation functions, for which equilibrium values and the correlation functions themselves may be rapidly calculated as discussed later in Appendix 2.4.7.

We proceed with the remaining correlation functions for motion of the interaction within frame $f$, first calculating $C_{01}^{v-f}(t)$ and $C_{0-1}^{v-f}(t)$.

$$C_{01}^{v-f}(t) = \left\langle D_{01}^2(\Omega_{\tau,t+\tau}^{v-f}) \right\rangle_\tau = \left\langle \exp(-i\gamma_{\tau,t+\tau}^{v-f})\sqrt{\frac{3}{8}}\sin(2\beta_{\tau,t+\tau}^{v-f}) \right\rangle_\tau$$

$$= \sqrt{\frac{3}{2}} \left\langle \left(\cos\gamma_{\tau,t+\tau}^{v-f}\sin\beta_{\tau,t+\tau}^{v-f}\right)\cos\beta_{\tau,t+\tau}^{v-f} \right\rangle_\tau - i\sqrt{\frac{3}{2}} \left\langle \left(\sin\gamma_{\tau,t+\tau}^{v-f}\sin\beta_{\tau,t+\tau}^{v-f}\right)\cos\beta_{\tau,t+\tau}^{v-f} \right\rangle_\tau$$

$$= \sqrt{\frac{3}{2}} \left\langle \left(\sum_{\zeta=x,y,z}\zeta_X^{-f}(\tau)\zeta_Z^{-f}(t+\tau)\right)\left(\sum_{\zeta=x,y,z}\zeta_Z^{-f}(\tau)\zeta_Z^{-f}(t+\tau)\right) \right\rangle_\tau$$

$$-i\sqrt{\frac{3}{2}} \left\langle \left(\sum_{\zeta=x,y,z}\zeta_Y^{-f}(\tau)\zeta_Z^{-f}(t+\tau)\right)\left(\sum_{\zeta=x,y,z}\zeta_Z^{-f}(\tau)\zeta_Z^{-f}(t+\tau)\right) \right\rangle_\tau \qquad (A7)$$

$$= \sqrt{\frac{3}{2}} \sum_{\zeta=x,y,z}\sum_{\xi=x,y,z} \left\langle \zeta_X^{-f}(\tau)\xi_Z^{-f}(\tau)\zeta_Z^{-f}(t+\tau)\xi_Z^{-f}(t+\tau) \right\rangle_\tau - i\left\langle \zeta_Y^{-f}(\tau)\xi_Z^{-f}(\tau)\zeta_Z^{-f}(t+\tau)\xi_Z^{-f}(t+\tau) \right\rangle_\tau$$

$$C_{0-1}^{v-f}(t) = -\left(C_{01}^{v-f}(t)\right)^*$$

Continuing, we calculate $C_{02}^{v-f}(t)$ and $C_{0-2}^{v-f}(t)$.

$$C_{02}^{v-f}(\tau) = \left\langle D_{02}^2(\Omega_{\tau,t+\tau}^{v-f})\right\rangle_\tau = \left\langle \exp(-2i\gamma_{\tau,t+\tau}^{v-f})\sqrt{\frac{3}{8}}\sin^2\beta_{\tau,t+\tau}^{v-f}\right\rangle_\tau$$

$$= \sqrt{\frac{3}{8}}\left\langle \left(\cos\gamma_{\tau,t+\tau}^{v-f}\sin\beta_{\tau,t+\tau}^{v-f}\right)^2 - \left(\sin\gamma_{\tau,t+\tau}^{v-f}\sin\beta_{\tau,t+\tau}^{v-f}\right)^2\right\rangle_\tau$$

$$-i\sqrt{\frac{3}{2}}\left\langle \left(\cos\gamma_{\tau,t+\tau}^{v-f}\sin\beta_{\tau,t+\tau}^{v-f}\right)\left(\sin\gamma_{\tau,t+\tau}^{v-f}\sin\beta_{\tau,t+\tau}^{v-f}\right)\right\rangle_\tau$$

$$= \sqrt{\frac{3}{8}}\left\langle \left(\sum_{\zeta=x,y,z}\zeta_X^{-f}(\tau)\zeta_Z^{-f}(t+\tau)\right)^2 - \left(\sum_{\zeta=x,y,z}\zeta_Y^{-f}(\tau)\zeta_Z^{-f}(t+\tau)\right)^2\right\rangle_\tau$$

$$-i\sqrt{\frac{3}{2}}\left\langle \left(\sum_{\zeta=x,y,z}\zeta_X^{-f}(\tau)\zeta_Z^{-f}(t+\tau)\right)\left(\sum_{\zeta=x,y,z}\zeta_Y^{-f}(\tau)\zeta_Z^{-f}(t+\tau)\right)\right\rangle_\tau$$

$$= \sqrt{\frac{3}{2}}\sum_{\zeta=x,y,z}\sum_{\xi=x,y,z}\left[\left\langle \zeta_X^{-f}(\tau)\xi_X^{-f}(\tau)\zeta_Z^{-f}(t+\tau)\xi_Z^{-f}(t+\tau)\right\rangle_\tau - \left\langle \zeta_Y^{-f}(\tau)\xi_Y^{-f}(\tau)\zeta_Z^{-f}(t+\tau)\xi_Z^{-f}(t+\tau)\right\rangle_\tau\right.$$

$$\left. -2i\left\langle \zeta_X^{-f}(\tau)\xi_Y^{-f}(\tau)\zeta_Z^{-f}(t+\tau)\xi_Z^{-f}(t+\tau)\right\rangle_\tau\right]$$

$$C_{0-2}^{v-f}(\tau) = \left(C_{02}^{v-f}(\tau)\right)^*$$

. (A8)

.

## 2.2 Correlation functions for motion of frame *f* in *F*

### 2.2.1 Required Euler angles

In order to calculate the correlation function for motion of frame *f* in *F*, we must expand the terms in eqs. (45) and (46). Note that the rotation matrix may be replaced by a projection into frame *F*. To do this, we need *x*-, *y*-, and *z*-axes of frame *F* represented in frame *f*, that is terms $v_\zeta^{F-f}(\tau)$. These are given by

$$v_X^{F-f}(\tau) = \mathbf{R}_{ZYZ}^{-1}(\Omega_\tau^{f,F})\cdot[1,0,0]'$$
$$v_Y^{F-f}(\tau) = \mathbf{R}_{ZYZ}^{-1}(\Omega_\tau^{f,F})\cdot[0,1,0]'$$
$$v_Z^{F-f}(\tau) = \mathbf{R}_{ZYZ}^{-1}(\Omega_\tau^{f,F})\cdot[0,0,1]'$$

$$\mathbf{R}_{ZYZ}(\Omega_\tau^{f,F}) = \begin{pmatrix}(v_X^{F-f}(\tau))'\\(v_Y^{F-f}(\tau))'\\(v_Z^{F-f}(\tau))'\end{pmatrix}$$

. (A9)

Then, the terms in eqs. (45) and (46) can be expanded as

$$\cos\beta^{v:f-F}_{\tau,t+\tau} = \mathbf{v}^{-F}_Z(\tau) \cdot \left(\mathbf{R}_{ZYZ}(\Omega^{f-F}_{t+\tau}) \cdot \mathbf{v}^{-f}_Z(\tau)\right)$$

$$= \begin{bmatrix} x^{-F}_Z(\tau) & y^{-F}_Z(\tau) & z^{-F}_Z(\tau) \end{bmatrix} \cdot \begin{bmatrix} x^{F-f}_X(t+\tau) & y^{F-f}_X(t+\tau) & z^{F-f}_X(t+\tau) \\ x^{F-f}_Y(t+\tau) & y^{F-f}_Y(t+\tau) & z^{F-f}_Y(t+\tau) \\ x^{F-f}_Z(t+\tau) & y^{F-f}_Z(t+\tau) & z^{F-f}_Z(t+\tau) \end{bmatrix} \cdot \begin{bmatrix} x^{-f}_Z(\tau) \\ y^{-f}_Z(\tau) \\ z^{-f}_Z(\tau) \end{bmatrix}$$

$$= \sum_{\zeta=x,y,z} \zeta^{-F}_Z(\tau) \sum_{\lambda=x,y,z} \lambda^{F-f}_\zeta(t+\tau)\lambda^{-f}_Z(\tau)$$

$$\cos\alpha^{v:f-F}_{\tau,t+\tau}\sin\beta^{v:f-F}_{\tau,t+\tau} = -\sum_{\zeta=x,y,z} \zeta^{-F}_X(t+\tau) \sum_{\lambda=x,y,z} \lambda^{F-f}_\zeta(\tau)\lambda^{-f}_Z(t+\tau) \quad , \tag{A10}$$

$$\sin\alpha^{v:f-F}_{\tau,t+\tau}\sin\beta^{v:f-F}_{\tau,t+\tau} = \sum_{\zeta=x,y,z} \zeta^{-F}_Y(t+\tau) \sum_{\lambda=x,y,z} \lambda^{F-f}_\zeta(\tau)\lambda^{-f}_Z(t+\tau)$$

$$\sin\beta^{v:f-F}_{\tau,t+\tau}\cos\gamma^{v:f-F}_{\tau,t+\tau} = \sum_{\zeta=x,y,z} \zeta^{-F}_X(\tau) \sum_{\lambda=x,y,z} \lambda^{F-f}_\zeta(t+\tau)\lambda^{-f}_Z(\tau)$$

$$\sin\beta^{v:f-F}_{\tau,t+\tau}\sin\gamma^{v:f-F}_{\tau,t+\tau} = \sum_{\zeta=x,y,z} \zeta^{-F}_Y(\tau) \sum_{\lambda=x,y,z} \lambda^{F-f}_\zeta(t+\tau)\lambda^{-f}_Z(\tau)$$

where terms $\zeta^{-F}_Z(t+\tau)$ are selected out of $\mathbf{v}^{-F}_Z(\tau) = [x^{-F}_Z(\tau), y^{-F}_Z(\tau), z^{-F}_Z(\tau)]$, terms $\lambda^{F-f}_\zeta(\tau)$ are selected out of $\mathbf{v}^{F-f}_\zeta(\tau) = [x^{F-f}_\zeta(\tau), y^{F-f}_\zeta(\tau), z^{F-f}_\zeta(\tau)]$, and terms $\lambda^{-f}_Z(\tau)$ are selected out of $\mathbf{v}^{-f}_Z(\tau) = [x^{-f}_Z(\tau), y^{-f}_Z(\tau), z^{-f}_Z(\tau)]$.

*2.2.2 Calculating the correlation functions*

Next, we calculate the terms required for correlation functions due to motion of the frame *f*. We start with $C^{v:f-F}_{00}(t) = \left\langle D^2_{00}(\Omega^{v:f-F}_{\tau,t+\tau}) \right\rangle_\tau$.

$$C^{v:f-F}_{00}(\tau) = \left\langle D^2_{00}(\Omega^{v-f}_{\tau,t+\tau}) \right\rangle_\tau = \left\langle -\frac{1}{2} + \frac{3}{2}\cos^2\beta^{v:f-F}_{\tau,t+\tau} \right\rangle_\tau$$

$$= -\frac{1}{2} + \frac{3}{2}\left\langle \left(\sum_{\zeta=x,y,z} \zeta^{-F}_Z(\tau) \sum_{\lambda=x,y,z} \lambda^{F-f}_\zeta(t+\tau)\lambda^{-f}_Z(\tau)\right)^2 \right\rangle_\tau \tag{A11}$$

$$= -\frac{1}{2} + \frac{3}{2}\sum_{\zeta,\xi,\lambda,\eta=x,y,z} \left\langle \zeta^{-F}_Z(\tau)\lambda^{-f}_Z(\tau)\xi^{-F}_Z(\tau)\eta^{-f}_Z(\tau)\lambda^{F-f}_\zeta(t+\tau)\eta^{F-f}_\xi(t+\tau) \right\rangle_\tau$$

Note that up until the insertion of the expression for $\cos\beta^{v:f-F}_{\tau,t+\tau}$, which is found in eq. (A10)), we follow the same steps as were found in eq. (A6). Then, we will omit these steps in most of the subsequent calculations. For completeness, we next calculate $C^{v:f-F}_{01}(t)$ and $C^{v:f-F}_{0-1}(t)$, although we note that these terms are not required for obtaining $C^{v:f-F}(t)$.

$$C_{01}^{v:f-F}(t) = \sqrt{\frac{3}{2}} \left\langle \left( \sum_{\zeta=x,y,z} \zeta_X^{-F}(\tau) \sum_{\lambda=x,y,z} \lambda_\zeta^{F-f}(t+\tau) \lambda_Z^{-f}(\tau) \right) \left( \sum_{\zeta=x,y,z} \zeta_Z^{-F}(\tau) \sum_{\lambda=x,y,z} \lambda_\zeta^{F-f}(t+\tau) \lambda_Z^{-f}(\tau) \right) \right\rangle_\tau$$

$$-i\sqrt{\frac{3}{2}} \left\langle \left( \sum_{\zeta=x,y,z} \zeta_Y^{-F}(\tau) \sum_{\lambda=x,y,z} \lambda_\zeta^{F-f}(t+\tau) \lambda_Z^{-f}(\tau) \right) \left( \sum_{\zeta=x,y,z} \zeta_Z^{-F}(\tau) \sum_{\lambda=x,y,z} \lambda_\zeta^{F-f}(t+\tau) \lambda_Z^{-f}(\tau) \right) \right\rangle_\tau$$

$$= \sqrt{\frac{3}{2}} \sum_{\zeta,\xi,\lambda,\eta=x,y,z} \left\langle \zeta_X^{-F}(\tau) \xi_Z^{-F}(\tau) \lambda_Z^{-f}(\tau) \eta_Z^{-f}(\tau) \lambda_\zeta^{F-f}(t+\tau) \eta_\xi^{F-f}(t+\tau) \right\rangle_\tau \quad \text{(A12)}$$

$$-i\sqrt{\frac{3}{2}} \sum_{\zeta,\xi,\lambda,\eta=x,y,z} \left\langle \zeta_Y^{-F}(\tau) \xi_Z^{-F}(\tau) \lambda_Z^{-f}(\tau) \eta_Z^{-f}(\tau) \lambda_\zeta^{F-f}(t+\tau) \eta_\xi^{F-f}(t+\tau) \right\rangle_\tau$$

$$C_{0-1}^{v:f-F}(t) = -(C_{01}^{v:f-F}(t))^*$$

Next, we calculate $C_{10}^{v:f-F}(t)$ and $C_{-10}^{v:f-F}(t)$, which appear in $C^{v:f-F}(t)$ (eq. (22)). We show all steps to see the influence of sign changes on the various terms.

$$C_{10}^{v:f-F}(t) = \left\langle D_{10}^2(\Omega_{\tau,t+\tau}^{v-f}) \right\rangle_\tau = \left\langle -\exp(-i\alpha_{\tau,t+\tau}^{v:f-F}) \sqrt{\frac{3}{8}} \sin(2\beta_{\tau,t+\tau}^{v:f-F}) \right\rangle$$

$$= -\sqrt{\frac{3}{2}} \left\langle \left( \cos\alpha_{\tau,t+\tau}^{v:f-F} \sin\beta_{\tau,t+\tau}^{v:f-F} \right) \cos\beta_{\tau,t+\tau}^{v:f-F} \right\rangle_\tau + i\sqrt{\frac{3}{2}} \left\langle \left( \sin\alpha_{\tau,t+\tau}^{v:f-F} \sin\beta_{\tau,t+\tau}^{v:f-F} \right) \cos\beta_{\tau,t+\tau}^{v:f-F} \right\rangle_\tau$$

$$= \sqrt{\frac{3}{2}} \left\langle \left( \sum_{\zeta=x,y,z} \zeta_X^{-F}(t+\tau) \sum_{\lambda=x,y,z} \lambda_\zeta^{F-f}(\tau) \lambda_Z^{-f}(t+\tau) \right) \left( \sum_{\zeta=x,y,z} \zeta_Z^{-F}(t+\tau) \sum_{\lambda=x,y,z} \lambda_\zeta^{F-f}(\tau) \lambda_Z^{-f}(t+\tau) \right) \right\rangle_\tau$$

$$+i\sqrt{\frac{3}{2}} \left\langle \left( \sum_{\zeta=x,y,z} \zeta_Y^{-F}(t+\tau) \sum_{\lambda=x,y,z} \lambda_\zeta^{F-f}(\tau) \lambda_Z^{-f}(t+\tau) \right) \left( \sum_{\zeta=x,y,z} \zeta_Z^{-F}(t+\tau) \sum_{\lambda=x,y,z} \lambda_\zeta^{F-f}(\tau) \lambda_Z^{-f}(t+\tau) \right) \right\rangle_\tau \quad \text{(A13)}$$

$$= \sqrt{\frac{3}{2}} \sum_{\zeta,\xi,\lambda,\eta=x,y,z} \left\langle \zeta_X^{-F}(t+\tau) \xi_Z^{-F}(t+\tau) \lambda_Z^{-f}(t+\tau) \eta_Z^{-f}(t+\tau) \lambda_\zeta^{F-f}(\tau) \eta_\xi^{F-f}(\tau) \right\rangle_\tau$$

$$+i\sqrt{\frac{3}{2}} \sum_{\zeta,\xi,\lambda,\eta=x,y,z} \left\langle \zeta_Y^{-F}(t+\tau) \xi_Z^{-F}(t+\tau) \lambda_Z^{-f}(t+\tau) \eta_Z^{-f}(t+\tau) \lambda_\zeta^{F-f}(\tau) \eta_\xi^{F-f}(\tau) \right\rangle_\tau$$

$$C_{-10}^{v:f-F}(t) = -(C_{10}^{v:f-F}(t))^*$$

We compare the result for $C_{01}^{v:f-F}(t)$ and $C_{10}^{v:f-F}(t)$. We note that the results are fairly similar, excepting a sign change on the imaginary part, and the times $\tau$ and $t+\tau$ are swapped. If we assume symmetry of the correlation functions in time (i.e. $\langle p(\tau)q(t+\tau)\rangle_\tau = \langle p(\tau)q(-t+\tau)\rangle_\tau$) we obtain

$$\begin{aligned}\langle p(\tau)q(t+\tau)\rangle_\tau &= \langle p(\tau)q(-t+\tau)\rangle_\tau \\ \text{shift } \tau \text{ to } t+\tau & \\ &= \langle p(t+\tau)q(-t+t+\tau)\rangle_\tau = \langle p(t+\tau)q(\tau)\rangle_\tau\end{aligned} \quad \text{(A14)}$$

We have replaced $\tau$ with $t+\tau$ in the last line, where the average over all times $\tau$ allows us to make this shift without changing the outcome of the average. Then, if the correlation functions are symmetric in time, we may swap all the times $\tau$ and $t+\tau$. Note that the assumption of time symmetry could break down, for example, if we have a rotation that moves preferentially clockwise or counterclockwise, although this would be unusual. Then, we obtain

$$C_{10}^{v:f-F}(t) = \sqrt{\frac{3}{2}} \sum_{\zeta,\xi,\lambda,\eta=x,y,z} \left\langle \zeta_X^{-F}(\tau)\xi_Z^{-F}(\tau)\lambda_Z^{-f}(\tau)\eta_Z^{-f}(\tau)\lambda_\zeta^{F-f}(t+\tau)\eta_\xi^{F-f}(t+\tau)\right\rangle_\tau$$
$$+i\sqrt{\frac{3}{2}} \sum_{\zeta,\xi,\lambda,\eta=x,y,z} \left\langle \zeta_Y^{-F}(\tau)\xi_Z^{-F}(\tau)\lambda_Z^{-f}(\tau)\eta_Z^{-f}(\tau)\lambda_\zeta^{F-f}(t+\tau)\eta_\xi^{F-f}(t+\tau)\right\rangle_\tau$$
(A15)

The result is the complex conjugate of the result in eq. (A12). We can see that this occurs due to first, an overall sign change for $d_{01}^2(\beta_{\tau,t+tau}^{v:f-F})$ vs. $d_{10}^2(\beta_{\tau,t+tau}^{v:f-F})$ (see Table 2). However, the real part includes a second sign swap, due to the sign difference for the terms found in eq. (A10).

Continuing, we calculate $C_{02}^{v:f-F}(t)$ and $C_{0-2}^{v:f-F}(t)$ (again, these terms are not required for calculating $C^{v:f-F}(t)$, we only include for completeness).

$$C_{02}^{v:f-F}(t)$$
$$=\sqrt{\frac{3}{2}}\left\langle \left(\sum_{\zeta=x,y,z}\zeta_X^{-F}(\tau)\sum_{\lambda=x,y,z}\lambda_\zeta^{F-f}(t+\tau)\lambda_Z^{-f}(\tau)\right)^2 - \left(\sum_{\zeta=x,y,z}\zeta_Y^{-F}(\tau)\sum_{\lambda=x,y,z}\lambda_\zeta^{F-f}(t+\tau)\lambda_Z^{-f}(\tau)\right)^2 \right\rangle_\tau$$
$$-i\sqrt{\frac{3}{2}}\left\langle \left(\sum_{\zeta=x,y,z}\zeta_X^{-F}(\tau)\sum_{\lambda=x,y,z}\lambda_\zeta^{F-f}(t+\tau)\lambda_Z^{-f}(\tau)\right)\left(\sum_{\zeta=x,y,z}\zeta_Y^{-F}(\tau)\sum_{\lambda=x,y,z}\lambda_\zeta^{F-f}(t+\tau)\lambda_Z^{-f}(\tau)\right)\right\rangle_\tau$$
$$=\sqrt{\frac{3}{2}}\sum_{\zeta,\xi,\lambda,\eta=x,y,z}\left[\left\langle \zeta_X^{-F}(\tau)\xi_X^{-F}(\tau)\lambda_Z^{-f}(\tau)\eta_Z^{-f}(\tau)\lambda_\zeta^{F-f}(t+\tau)\eta_\xi^{F-f}(t+\tau)\right\rangle_\tau\right.$$
$$\left.-\left\langle \zeta_Y^{-F}(\tau)\xi_Y^{-F}(\tau)\lambda_Z^{-f}(\tau)\eta_Z^{-f}(\tau)\lambda_\zeta^{F-f}(t+\tau)\eta_\xi^{F-f}(t+\tau)\right\rangle_\tau\right]$$
$$-i\sqrt{\frac{3}{2}}\sum_{\zeta,\xi,\lambda,\eta=x,y,z}\left\langle \zeta_X^{-F}(\tau)\xi_Y^{-F}(\tau)\lambda_Z^{-f}(\tau)\eta_Z^{-f}(\tau)\lambda_\zeta^{F-f}(t+\tau)\eta_\xi^{F-f}(t+\tau)\right\rangle_\tau$$
(A16)

$$C_{0-2}^{v:f-F}(t) = (C_{02}^{v:f-F}(t))^*$$

Finally, we calculate $C_{20}^{v:f-F}(t)$ and $C_{-20}^{v:f-F}(t)$. Assuming symmetry of the correlation functions, we see that the only change in $C_{20}^{v:f-F}(t)$ relative to $C_{02}^{v:f-F}(t)$ is the sign of the expression for $\cos\alpha_{\tau,t+\tau}^{v:f-F}\sin\beta_{\tau,t+\tau}^{v:f-F}$ vs. $\sin\beta_{\tau,t+\tau}^{v:f-F}\cos\gamma_{\tau,t+\tau}^{v:f-F}$. This term appears in both the real and imaginary parts of the correlation functions. However, in the real part, the term is squared, resulting overall in $C_{20}^{v:f-F}(t)$ being the complex conjugate of $C_{02}^{v:f-F}(t)$.

$$C_{20}^{v:f-F}(t) = \sqrt{\frac{3}{2}} \sum_{\zeta,\xi,\lambda,\eta=x,y,z} \left[ \left\langle \zeta_X^{-F}(\tau)\xi_X^{-F}(\tau)\lambda_Z^{-f}(\tau)\eta_Z^{-f}(\tau)\lambda_\zeta^{F-f}(t+\tau)\eta_\xi^{F-f}(t+\tau) \right\rangle_\tau \right.$$
$$\left. - \left\langle \zeta_Y^{-F}(\tau)\xi_Y^{-F}(\tau)\lambda_Z^{-f}(\tau)\eta_Z^{-f}(\tau)\lambda_\zeta^{F-f}(t+\tau)\eta_\xi^{F-f}(t+\tau) \right\rangle_\tau \right]$$
$$+ i\sqrt{\frac{3}{2}} \sum_{\zeta,\xi,\lambda,\eta=x,y,z} \left\langle \zeta_X^{-F}(\tau)\xi_Y^{-F}(\tau)\lambda_Z^{-f}(\tau)\eta_Z^{-f}(\tau)\lambda_\zeta^{F-f}(t+\tau)\eta_\xi^{F-f}(t+\tau) \right\rangle_\tau$$

$$C_{-20}^{v:f-F}(t) = (C_{20}^{v:f-F}(t))^*$$

(A17)

We summarize the relationships between the various correlation functions, which may be used to reduce computational time:

$$C_{01}^{v:f-F}(t) = -(C_{0-1}^{v:f-F}(t))^* = (C_{10}^{v:f-F}(t))^* = -C_{-10}^{v:f-F}(t)$$
$$C_{02}^{v:f-F}(t) = (C_{0-2}^{v:f-F}(t))^* = (C_{20}^{v:f-F}(t))^* = C_{-20}^{v:f-F}(t)$$

(A18)

## 2.3 Correlation functions for motion of frame *f* in *F* with a symmetry axis

### 2.3.1 Required Euler angles

We only require the cosine of the angles $\beta_{\tau,t+\tau}^{\text{sym}}$, obtained from the dot product of $\mathbf{v}_Z^{\text{sym}}(\tau)$ and $\mathbf{v}_Z^{\text{sym}}(t+\tau)$.

$$\cos\beta_{\tau,t+\tau}^{\text{sym}} = \sum_{\zeta=x,y,z} \zeta_Z^{\text{sym}}(\tau)\zeta_Z^{\text{sym}}(t+\tau).$$

(A19)

The terms $\zeta_Z^{\text{sym}}(\tau)$ are selected out of $\mathbf{v}_Z^{\text{sym}}(\tau) = [x_Z^{\text{sym}}(\tau), y_Z^{\text{sym}}(\tau), z_Z^{\text{sym}}(\tau)]$.

### 2.3.2 Calculating the correlation function

Following the same steps as in eq. (A6), we may then obtain the correlation function

$$C^{v-f}(t) = \left\langle D_{00}^2(\Omega_{\tau,t+\tau}^{\text{sym}}) \right\rangle_\tau - \frac{1}{2} + \frac{3}{2} \sum_{\zeta=x,y,z}\sum_{\xi=x,y,z} \left\langle \zeta_Z^{\text{sym}}(\tau)\xi_Z^{\text{sym}}(\tau)\zeta_Z^{\text{sym}}(t+\tau)\xi_Z^{\text{sym}}(t+\tau) \right\rangle_\tau.$$

(A20)

## 3 Accelerating Calculations

In eqs. (A6)-(A20), we find that the various correlation functions can always be expressed as a sum over terms of the form $\left\langle f(\tau)g(t+\tau) \right\rangle_\tau$ (sometimes with a constant offset, in the case of $C_{00}^{v-f}(t)$ and $C_{00}^{v:f-F}(t)$). For a given element of the summations, $f(\tau)$ is the product of all terms that are a function of $\tau$, and $g(t+\tau)$ is the product of all terms that are a function of $t+\tau$. Then, each term of the summation is itself a linear correlation function, for which we may quickly obtain its equilibrium value, i.e. $\lim_{t\to\infty}\left\langle f(\tau)g(t+\tau) \right\rangle_\tau$, and the correlation function itself. We first discuss how to obtain the equilibrium values.

## 3.1 Rapid calculation of equilibrium values of linear correlation functions

The contribution to the equilibrium value of each term, that is $\lim_{t\to\infty}\langle f(\tau)g(t+\tau)\rangle_\tau$, is obtained by noting that $f(\tau)$ and $g(t+\tau)$ must be uncorrelated if they are separated by infinite time. Then, we obtain

$$\lim_{t\to\infty}\langle f(\tau)g(t+\tau)\rangle_\tau = \langle f(\tau)\rangle_\tau \langle g(t+\tau)\rangle_\tau$$
$$= \langle f(\tau)\rangle_\tau \langle g(\tau)\rangle_\tau \tag{A21}$$

In the first step, we separate the terms since they are uncorrelated, and in the latter step, we note that the time shift makes no difference due to the average over all $\tau$. As a quick example, we find $S^2$ for some motion within frame *f*.

$$S^2 = \lim_{t\to\infty}\langle D_{00}^2(\Omega_{\tau,t+\tau}^{v-f})\rangle_\tau = -\frac{1}{2} + \frac{3}{2}\sum_{\zeta=x,y,z}\sum_{\xi=x,y,z}\langle \zeta_z^{-f}(\tau)\xi_z^{-f}(\tau)\zeta_z^{-f}(t+\tau)\xi_z^{-f}(t+\tau)\rangle_\tau$$
$$= -\frac{1}{2} + \frac{3}{2}\sum_{\zeta=x,y,z}\sum_{\xi=x,y,z}\langle \zeta_z^{-f}(\tau)\xi_z^{-f}(\tau)\rangle_\tau \langle \zeta_z^{-f}(t+\tau)\xi_z^{-f}(t+\tau)\rangle_\tau \tag{A22}$$
$$= -\frac{1}{2} + \frac{3}{2}\sum_{\zeta=x,y,z}\sum_{\xi=x,y,z}\langle \zeta_z^{-f}(\tau)\xi_z^{-f}(\tau)\rangle_\tau^2$$

This is the familiar result for deriving $S^2$ [53–55], given originally in a somewhat different form by Henry and Szabo [56] (see eqs. 8 and 11b). Note that in this approach, there is an inherent assumption that the trajectory is equilibrated, i.e. it is a good representation of the thermal equilibrium distribution of configurations of the system. Caution should be taken when working with trajectories for which this is not the case; slow motion will frequently lead to an un-equilibrated trajectory.

## 3.2 Rapid calculation of linear correlation functions

Next, we show how to rapidly obtain the contribution of each term, $\langle f(\tau)g(t+\tau)\rangle_\tau$, to the total correlation function. Note that this term can be calculated explicitly, via summation of all pairs of time points separated by *t*:

$$\langle f(\tau)g(t_n+\tau)\rangle_\tau = \frac{1}{N-n}\sum_{i=0}^{N-n-1} f(t_i)g(t_{i+n}) \tag{A23}$$

Here, we index *N* time points from 0 to *N*–1, so that the $n^{th}$ time point of the correlation function is calculated from all pairs of time points separated by *n* time points. This calculation can be very slow, especially in the case of long trajectories. However, each term, $\langle f(\tau)g(t+\tau)\rangle_\tau$ is itself a linear correlation function, allowing us to apply the discrete convolution theorem to obtain the correlation function quickly via Fourier transform. Then, for two *periodic* functions (with period *N*), the correlation can be calculated as:

$$\langle f(\tau)g(t+\tau)\rangle_\tau = \frac{1}{N}\text{IDFT}\left(\text{DFT}(f(\tau))^* \times \text{DFT}(g(\tau))\right) \tag{A24}$$

if $f(\tau)$ and $g(\tau)$ are periodic

Here, DFT and IDFT are the discrete Fourier transform and its inverse, respectively, where we take the complex conjugate of the DFT of $f(\tau)$. $N$ is the number of time points. While in principle the IDFT may return complex values, this function will always result in a real correlation function if $f(\tau)$ and $g(\tau)$ are real (within numerical error– practically one should extract only the real part to remove any residual imaginary component).

However, this is valid for two periodic functions, that is, if we have $f(\tau)$ which has period $N$, then for each time point $f(\tau_n) = f(\tau_{n+N})$. Therefore, eq. (A24) actually returns

$$\frac{1}{N}\sum_{i=0}^{N-1} f(t_i)g(t_{\text{mod}(i+n,N)}), \tag{A25}$$

where each term is the result of summing over $N$ time points (instead of $N-n$, $n$ being the index of the current time point). The modulus causes values of $i+n \geq N$ to be replaced by $i+n-N$, so that the resulting correlation function includes pairing of time points resulting from the periodicity. This is incorrect for the desired correlation function since we do not have periodicity of the motion: all terms $i+n \geq N$ should be omitted or set to zero. A simple way to achieve this while still taking advantage of the speed of Fourier transforms is to zero-fill the two functions to twice the number of time points in the function, then $f^{ZF}(\tau)$ and $g^{ZF}(\tau)$ have $2N$ time points, where the last $N$ points are zero.

$$f^{ZF}(t_n) = \begin{cases} f(t_n) & \text{if } n < N \\ 0 & \text{otherwise} \end{cases}$$
$$g^{ZF}(t_n) = \begin{cases} g(t_n) & \text{if } n < N \\ 0 & \text{otherwise} \end{cases} \tag{A26}$$

Next, we apply the DFT to both $f^{ZF}(\tau)$ and $g^{ZF}(\tau)$, take the product, and apply the IDFT. We discard all time points $n \geq N$, so that our correlation function is the correct length, and renormalize the result.

for $n < N$:

$$\langle f(\tau)g(t_n+\tau)\rangle_\tau = \frac{1}{N-n}\left[\text{IDFT}\left(\text{DFT}(f^{ZF}(\tau))^* \times \text{DFT}(g^{ZF}(\tau))\right)\right]_n \tag{A27}$$

From eq. (A25) (where $N$ is now replaced by $2N$), we find for $n < N$ that non-zero elements of $f(t_i)$ are only paired with elements of $g(t_{i+n})$ for which $i+n < N$, or elements that are zero. Then, the $n^{\text{th}}$ time point is paired with $N$ points, but $n$ of those are zero. This leads to our normalization by a factor of $N-n$ as opposed to $N$, as was done in eq. (A24).

Practically, we calculate all time points simultaneously with the Fourier transform approach, and subsequently must apply a different normalization to each term of the result. We denote this by indicating that the formula is for the specific time point, $t_n$, using a subscript.

Note that when we calculate one of the full correlation functions in eqs. (A6)-(A20), we may perform the summation before applying the IDFT and normalization, since the IDFT is a linear operation acting on a sum. This saves considerable computational time, since it reduces the number of DFTs required by roughly 1/3 (however, the DFTs inside the summation must be performed before multiplication of the results of the DFTs).